\documentclass{aa}  

\usepackage{graphicx}
\usepackage{txfonts}
\usepackage{lscape}
\begin{document}

   \title{On the specific energy and pressure in near-Earth magnetic clouds}
  \titlerunning{Specific energy, total pressure in near-Earth magnetic clouds}
  \authorrunning{Bhattacharjee et al.}

   \author{Debesh Bhattacharjee
          \inst{1}          
          Prasad Subramanian
          \inst{1}
          Angelos Vourlidas
          \inst{2}
          Teresa Nieves-Chinchilla
          \inst{3}
          Niranjana Thejaswi
          \inst{4}
          \and
          Nishtha Sachdeva
          \inst{5}}

   \institute{Indian Institute of Science Education and Research, Pune
Dr. Homi Bhabha Road, Pashan, Pune 411008,
India\\
\and
The Johns Hopkins University Applied Physics Laboratory, Laurel MD, USA. \\
\and
Heliophysics science division, NASA-Goddard Space Flight Center, Greenbelt, MD (USA).
\\
\and
Department of Physics, The University of Arizona, Tucson, AZ 85721, USA \\              
         \and
             Department of Climate and Space Science and Engineering, University of Michigan, Ann Arbor, USA \\}


 
  \abstract
   {The pressure and energy density of the gas and magnetic field inside solar coronal mass ejections (in relation to that in the ambient solar wind) is thought to play an important role in determining their dynamics as they propagate through the heliosphere.}
   {We compare the specific energy (${\rm erg\,g^{-1}}$) [comprising kinetic ($H_{\rm k}$), thermal ($H_{\rm th }$) and magnetic field ($H_{\rm mag}$) contributions] inside MCs and the solar wind background.
   	We examine if the excess thermal + magnetic pressure and specific energy inside MCs (relative to the background) is correlated with their propagation and internal expansion speeds. We ask if the excess thermal + magnetic specific energy inside MCs might make them resemble rigid bodies in the context of aerodynamic drag.}
   {We use near-Earth {\em in-situ} data from the WIND spacecraft to identify a sample of 152 well observed interplanetary coronal mass ejections and their MC counterparts. We compute various metrics using these data to address our questions.}
   {We find that the total specific energy ($H$) inside MCs is approximately equal to that in the background solar wind. We find that the the excess (thermal + magnetic) pressure and specific energy are not well correlated with the near-Earth propagation and expansion speeds. We find that the excess thermal+magnetic specific energy $\gtrsim$ the specific kinetic energy of the solar wind incident on  81--89 \% of the MCs we study. This might explain how MCs retain their structural integrity and resist deformation by the solar wind bulk flow.}
   {}


   \maketitle
%

\section{Introduction}
Earth-directed Coronal mass ejections (CMEs) originating from the solar corona are the primary drivers of geomagnetic storms. Realistic estimates of Sun-Earth CME propagation times and arrival velocities are therefore an important component of space weather forecasting. Understanding the dynamics of CMEs and the forces leading to their propagation and expansion is crucial to this endeavor. Approaches to this problem range from early analytical models for the entire Sun-Earth propagation \citep{1996ChenJGR, 1996KumarRustJGR}, semi-analytical models that apply only to the aerodynamic drag-dominated phase of the propagation \citep{2004CargillSoPh, 15NApJ, 2013VrsSoPh} to detailed 3D MHD models \citep{1999LinkerJGR, 2009OdstrcilSoPh, 2020Keppens, 2012TothJCoPh}. Some efforts have focussed on characterizing the internal magnetic structure of the interplanetary counterparts of CMEs (ICMEs) \citep{1982KleinBurlagaJGR, 2016NCApJ} and others have focussed on comprehensive characterizations of ICME peoperties \citep{2010RichCaneSoPh, 2021TemmerLRSP, 2006ForsythSSRv}.

Despite these advances, there are still some fairly basic issues that remain to be addressed in this area. CME expansion provides a concrete window into some of these issues. It is well known that CMEs translate as well as expand as they travel through the heliosphere - CMEs are observed to expand in typical coronograph fields of view \citep{2000StJGR} and beyond \citep{2010LugazApJ,2009WebbSoPh}. CME expansion has also been confirmed using {\em in-situ} observations in the heliosphere \citep{1998BthAnGeo,2004WangJGRA} and near the Earth \citep{2007DassoSoPh}. The expansion is thought to occur because the interior of the CME is over-pressured with respect to its surroundings \citep[e.g.,][]{2006vonSteigerSSRv, 2019ScoliniA&A,2009DDA&A,2022Verbeke}, although some contend that the expansion is an outcome of CME magnetic field rearrangement \citep{1996KumarRustJGR}. Some authors \citep[e.g.,][]{2014GopalGeoRL,2015GopalJGRA,2022Dagnew},  ask if the abundance of halo CMEs during solar cycle 24 is because the ambient solar wind pressure is generally lower, leading CMEs to be more over-pressured (with respect to the surroundings) than usual. CME expansion speeds are also known to be lower than the Alfv\'en speeds in the ambient solar wind \citep{1982KleinBurlagaJGR,2020LugazApJ} - this is another instance of comparison between the CME plasma and that of the surrounding solar wind. CME identification using {\em in-situ} data also relies on a comparison between the CME and the ambient solar wind plasma - one of the well accepted criteria for identifying near-Earth magnetic clouds is that it is a low plasma beta structure, relative to the background solar wind \citep{1982KleinBurlagaJGR,2003LeppingSoPh}. 
Evidently, a comparison of the thermal and magnetic pressure inside CMEs (relative to the background solar wind) is an issue of considerable interest. In this paper, we use near-Earth {\em in-situ} data to compare the plasma inside a large sample of well-observed magnetic clouds (MCs) with respect to their surroundings. Besides comparing the thermal and magnetic pressures, we also compute the specific energy which is a conserved quantity in an ideal magnetized flow, and serves as a useful reference quantity. 
Such an exercise has not been carried out for a large sample of well-observed events to the best of our knowledge, and it allows us to reach a number of useful conclusions.

The data used in this study are described in \S~\ref{S - data} and the total specific energy for an ideal magnetized fluid is discussed in \S~\ref{S - Bernoulli}. The total specific energy inside the MC is compared with that in the ambient solar wind in \S~\ref{Sub - hmc_vs_hbg}, while the thermal+magnetic specific energy inside MCs is compared with that in the ambient solar wind in \S~\ref{Sub - total specific energy}. We compare the thermal+magnetic pressure inside MCs with that in the ambient solar wind in \S~\ref{Sub - total pressure}. The role of the excess thermal+magnetic specific energy and pressure in influencing MC propagation and expansion is discussed in \S~\ref{Sub - cxcpwithvexp}. We speculate how the excess thermal+magnetic specific energy could contribute to maintaining the structural integrity of MCs in \S~\ref{S - rigidity} and present the conclusions in \S~\ref{S - conclusions}.

\section{Data}
\label{S - data}
We use {\em in-situ} data from the WIND spacecraft (\url{https://wind.nasa.gov/}) for this study. The WIND ICME catalogue (\url{https://wind.nasa.gov/ICMEindex.php}) provides a sample of well observed Earth directed ICMEs as observed by the WIND spacecraft \citep{2019NCSoPh, 18NCSo} at 1 AU. In this paper, we limit our study to the analysis of magnetic clouds (MCs), which are the magnetically well-structured parts of ICMEs, with typically better defined boundaries and expansion speeds \citep{1982KleinBurlagaJGR}. The MCs associated with these ICMEs are classified into different categories depending upon how well the observed plasma parameters fit the expectations of a static flux rope configuration. Of the ICMEs observed between 1995 and 2015 listed on the WIND website, we first shortlist MCs that are categorized as F+ and Fr events. These events best fit the expectations of the flux rope model \citep{2016NCApJ, 18NCSo}. Fr events indicate MCs with a single magnetic field rotation between $90^{\circ}$ and $180^{\circ}$ and F+ events indicate MCs with a single magnetic field rotation greater than $180^{\circ}$. We further shortlist events that are neither preceded nor followed by any other ICMEs or ejecta within a window of two days ahead of and after the event under consideration. This helps us exclude possibly interacting events. Our final shortlist comprises 152 ICMEs, which are listed in Table \ref{S - Table A} of the appendix. Since we intend to compare the pressure and specific energy inside MCs with that in the ambient/background solar wind, we need concrete criteria to define the background. Ideally, the background should be quiet and should be in the vicinity of the MC. Accordingly, we use two different solar wind backgrounds for each event. The first kind of background, which we call BG1, is a 24-hour window within 5 days preceding the event that satisfies the following conditions: a) the rms fluctuations of the solar wind velocity for this 24-hour period should not exceed 10\% of the mean value b) the rms fluctuations of the total magnetic field for this 24-hour period should not exceed 20\% of the mean value c) there are no magnetic field rotations. We find that the average plasma beta in the background is at least 1.5 times higher than that in the MC. Criteria a) and b) ensure that the background is quiet. Criterion c) distinguishes between the background and the MC, since MCs are characterized by large rotations of magnetic field components and low plasma beta. The second background, which we call BG2, is a 24-hour period immediately preceding the event. We use the term `solar wind' only for the ambient/background solar wind throughout this study. 
\begin{figure}[h!]    
                                \centering
  \includegraphics[width=\hsize]{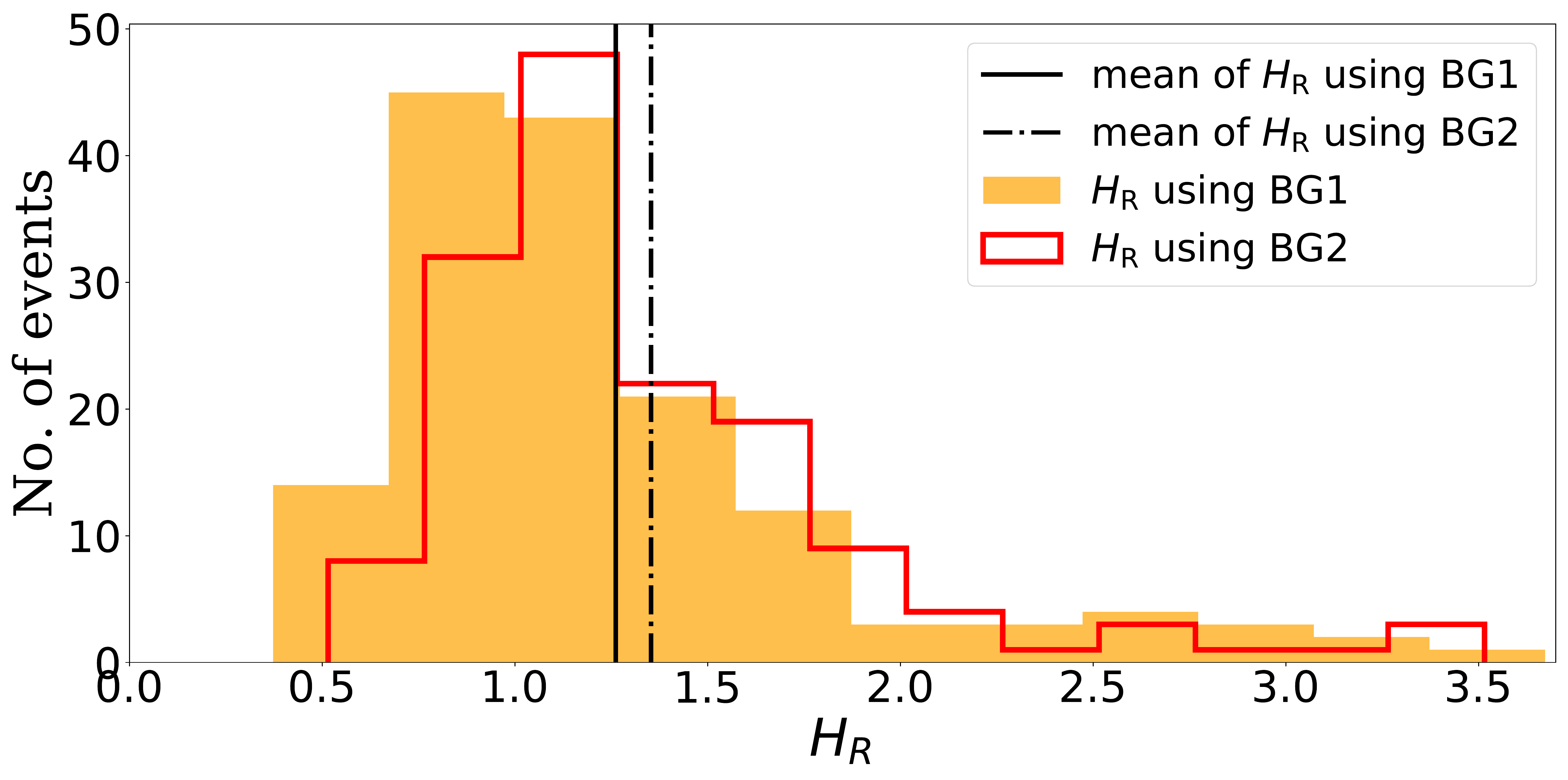}
  
  \includegraphics[width = \hsize]{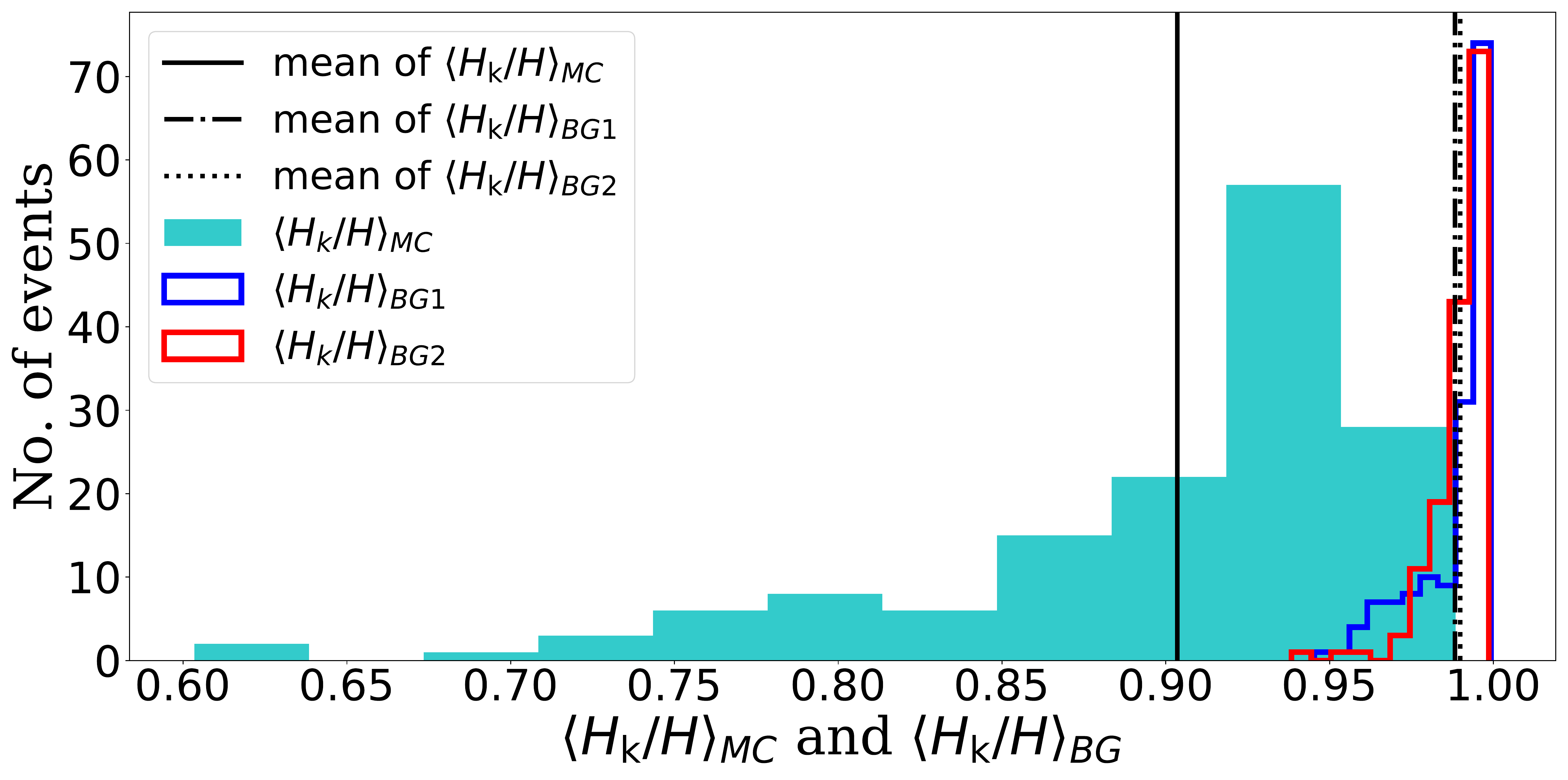} 

\caption{Histograms of $H_{\rm R}$ and $\langle H_{\rm k}/H \rangle$ for all the events listed in Table~\ref{S - Table A}. The top panel shows the histograms for $H_{\rm R}$ (Equation \ref{eq:HR}) with $\gamma = 5/3$ using two different backgrounds, BG1 and BG2. The mean, median and most probable value of $H_{\rm R}$ using BG1 are 1.26, 1.1 and 0.97 respectively. Using BG2, they are 1.36, 1.19 and 1.12 respectively. The histograms in the bottom panel display the ratio of $H_{\rm k}$ to $H$ inside the MC ($\langle H_{\rm k}/H \rangle_{MC}$) and in the backgrounds ($\langle H_{\rm k}/H \rangle_{BG1}$, $\langle H_{\rm k}/H \rangle_{BG2}$) using $\gamma = 5/3$. The mean, median and the most probable value of $\langle H_{\rm k}/H \rangle_{MC}$ are 0.90, 0.92 and 0.94 respectively. The mean, median and the most probable value of $\langle H_{k}/H \rangle_{BG1}$ are 0.98 , 0.99 and 0.995 respectively and those of $\langle H_{\rm k}/H \rangle_{BG2}$ are 0.99, 0.99 and 0.996 respectively. The mean value of each histogram is marked by a vertical line.}
   \label{Figure H}
   \end{figure}  
   
\section{The specific energy for an ideal magnetized fluid}
\label{S - Bernoulli}
In the lab frame, the conservative form of the ideal MHD energy equation is [\citep[Equation 65.10, Chapter 8, ][]{1987Landau}; \citep[Chapter 4 of ][]{2003BoydSanderson,2005Kulsrud}] 
\begin{equation}
\frac{\partial E}{\partial t} = - {\mathbf \nabla} \, . \, \left[\frac{\rho v^2}{2} \mathbf{v} + \frac{\gamma}{\gamma - 1}P_{\rm th} \mathbf{v} - \frac{(\mathbf{v} \times \mathbf{B})\times \mathbf{B}}{4\pi} \right]\, \, ,
\label{eq: consv}
\end{equation}
where $E$ is the energy density (${\rm erg\,cm^{-3}}$) of a parcel of fluid, ${\mathbf v}$ is the fluid velocity (${\rm cm\,s^{-1}}$), $\rho$ is the mass density of the fluid (${\rm g\,cm^{-3}}$), $P_{\rm th}$ is the thermal pressure (including contributions from protons and electrons), $\gamma$ is the polytropic index and ${\mathbf B}$ is the total magnetic field. The term inside the square brackets on the right hand side (RHS) of Equation~\ref{eq: consv} represents the total energy flux (${\rm erg\,cm^{-2}\,s^{-1}}$). In what follows, we will write the energy flux as $\rho v H\mathbf {\hat{p}}$, where $H$ is the total specific energy (${\rm erg\,g^{-1}}$) and $\mathbf{\hat{p}}$ is the unit vector directed along the total energy flux. 
$H$ contains contributions from the bulk motion of the fluid, the thermal energy and the magnetic field. The quantity 
\begin{equation}
H_{\rm k} \equiv (1/2) v^{2}
\label{eq: H_k}
\end{equation}
is the specific kinetic energy (${\rm erg\,g^{-1}}$) of the fluid due to its bulk motion and
\begin{equation}
H_{\rm th} \equiv \frac{\gamma}{\gamma - 1} \frac{P_{\rm th}}{\rho}
\label{eq: H_gas}
\end{equation}
is the specific thermal energy (${\rm erg\,g^{-1}}$) associated with the fluid. In order to understand the contribution to the specific energy from the magnetic field, we examine the Poynting flux ($\mathbf S$), which is the energy flux (${\rm erg\,cm^{-2}\,s^{-1}}$) carried by the electromagnetic field, and is defined by
 \begin{equation}
 	\begin{split} 
 {\mathbf S} \equiv (c/4 \pi) {\mathbf E} \times {\mathbf B} = (1/4 \pi) {\mathbf B} \times ({\mathbf v} \times {\mathbf B}) = \mathbf{(1 / 4 \pi) [\mathbf{v} B^2 - \mathbf{B}(\mathbf{v} \cdot \mathbf{B})]} \\ = 
 (1/4 \pi) \mathbf {v_{\perp}} B^{2} = (1/4 \pi) v_{\perp} B^{2} \mathbf {\hat{n}} \, ,
 \label{eq:poynting}
 \end{split} 
\end{equation}
where $\mathbf{v_\perp}$ and ${\mathbf v_{\parallel}}$ are the components of the fluid velocity perpendicular to and parallel to the magnetic field respectively and $\mathbf {\hat{n}}$ is the unit vector perpendicular to $\mathbf{B}$.
Equation~\ref{eq:poynting} assumes an infinitely conducting fluid and induction-only electric field. It follows from Equation~\ref{eq:poynting} that the energy density (${\rm erg\,cm^{-3}}$) associated with the magnetic field is $B^{2}/4 \pi$ \citep{09Pconf}. Accordingly, we define the specific energy (${\rm erg\,g^{-1}}$) associated with the magnetic field as
\begin{equation}
H_{\rm mag} \equiv (B^{2}/4 \pi \rho)
\label{eq: H_mag}
\end{equation}
The quantity in the square brackets in Eq~\ref{eq: consv} is then
{\begin{eqnarray}
		\begin{split} 
\nonumber
\rho (H_{\rm k} + H_{\rm th}) (v_{||} \hat{\mathbf b} + v_{\perp} \hat{\mathbf n}) + \rho H_{\rm mag} v_{\perp} \hat{\mathbf n} = \rho v_{\perp} (H_{\rm k} + H_{\rm th} + H_{\rm mag}) \hat{\mathbf n} \\ + \rho v_{||} (H_{\rm k} + H_{\rm th}) \hat{\mathbf b} 
= \rho v \hat{\mathbf p} [\cos^2\theta (H_{\rm k} + H_{\rm th})^2 + \\ \sin^2\theta (H_{\rm k} + H_{\rm th} + H_{\rm mag})^2]^{1/2} = \rho v \hat {\mathbf p} H \, ,
\label{eq: consv1}
\end{split} 
\end{eqnarray}} 
where ${\rm sin}\theta \equiv v_{\perp}/v$,  ${\rm cos}\theta \equiv v_{||}/v$, $\hat{\mathbf b}$ is the unit vector along ${\mathbf B}$ and $\hat{\mathbf p}$ is a unit vector in the direction (which is neither along the fluid streamlines nor along the magnetic field) along which the total energy flux is directed and 
\begin{equation}
H \equiv [\cos^2\theta (H_{\rm k} + H_{\rm th})^2 + \sin^2\theta (H_{\rm k} + H_{\rm th} + H_{\rm mag})^2]^{1/2}.
\label{eq: totH}
\end{equation} 
The angle $\alpha$ that $\hat{\mathbf p}$ makes with $\hat{\mathbf b}$ is given by
\begin{equation}
{\rm tan} \, \alpha = \frac{v_{\perp} (H_{\rm k} + H_{\rm th} + H_{\rm mag})}{v_{||} (H_{\rm k} + H_{\rm th})}
\label{eq: p}
\end{equation}
In steady state, the total energy is conserved in ideal MHD, which means the left hand side of Eq~\ref{eq: consv} is zero. Using Eq~\ref{eq: consv1}, this means that
${\mathbf \nabla} \, \cdot \, (\rho v \hat {\mathbf p} H) = H \, {\mathbf \nabla} \, \cdot \, (\rho v \hat {\mathbf p}) + (\rho v \hat {\mathbf p}) \cdot {\mathbf \nabla} H$ = 0. In steady state, mass conservation implies that ${\mathbf \nabla} \, \cdot \, (\rho v \hat {\mathbf p}) = 0$, which means that $\rho\, v\, \mathbf{\hat{p}}\cdot \nabla H = 0$. In turn, this means that ${\mathbf \nabla} H \perp \hat{\mathbf p}$ and $H$ (Eq~\ref{eq: totH}) is constant along $\hat{\mathbf p}$.
This is unlike an unmagnetized fluid, where the total specific energy (often referred to as the Bernoulli parameter) is equal to the sum of $H_{k}$ and $H_{\rm th}$ and is conserved along the velocity streamlines. The direction $\hat{\mathbf p}$ along which $H$ is conserved (in ideal MHD) need not coincide with the line of the spacecraft intercept. Therefore, we only compare the average value of $H$ inside MCs with its average value in the background solar wind (rather than regarding $H$ as a conserved quantity along the spacecraft line of intercept). This is in the same spirit in which the thermal and magnetic pressures inside CMEs are compared with those in the background solar wind.
Furthermore, $H$ does not include non-ideal effects such as losses due to viscous and resistive heating. However, the bulk solar wind is characterized by very high fluid and magnetic Reynolds numbers, which is why ideal MHD is usually considered to be an adequate description for characterizing the solar wind. Even the plasma inside magnetic clouds has very high Lundquist numbers \citep{2022Debesh}, justifying an ideal MHD treatment. Finally, the expression for $H$ (Eq~\ref{eq: totH}) uses a polytropic index $\gamma$. The appropriate value to use for $\gamma$ inside ICMEs or MCs is not clear. A value of 5/3 would imply that the ICME plasma is cooling adiabatically and needs to be heated to maintain its temperature \citep{1996KumarRustJGR},
whereas a value of 1.2 implies efficient thermal conduction to the interior of the ICME from the solar corona and little additional heating \citep{1996ChenJGR}. A recent study using Helios and Parker Solar Probe (PSP) data postulates a polytropic index ranging from 1.35 to 1.57 for solar wind protons and an index ranging from 1.21 to 1.29 for solar wind electrons \citep{2022Dakeyo}. Another study using PSP data \citep{2020NicolaouApJ} claims a polytropic index $\approx 5/3$ for the solar wind plasma. The polytropic index of the CME plasma is thought to range from 1.35 to 1.8 \citep{2018MishraApJ}. In view of this, we use two values for $\gamma$ (5/3 and 1.2) in our calculations. 
\begin{figure}
   \centering
   \includegraphics[width=\hsize]{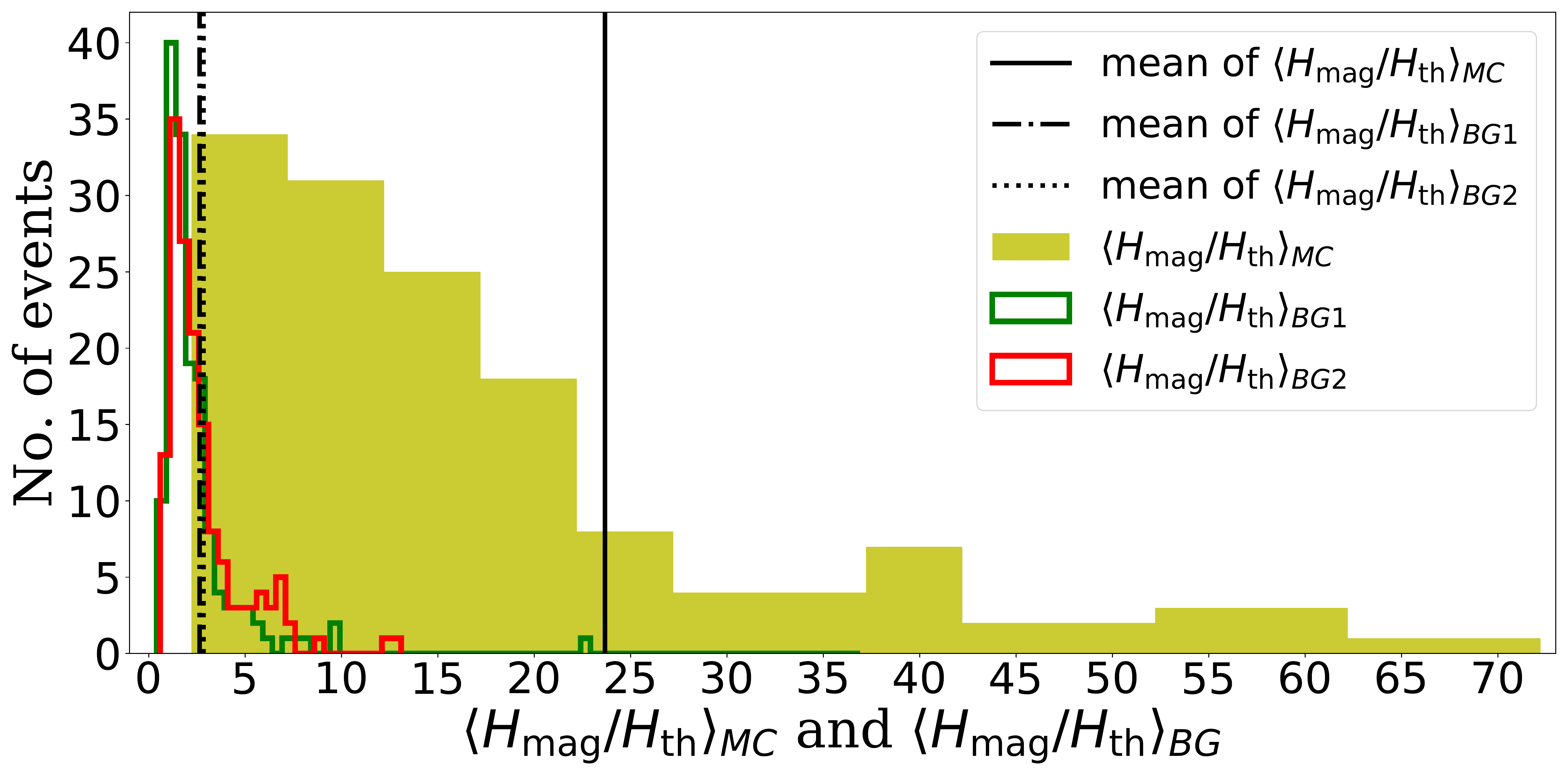}

      \caption{Histograms of $\langle H_{\rm mag} / H_{\rm th} \rangle$ (with $\gamma = 5/3$) inside the MC and in the two backgrounds, BG1 and BG2, for all the events listed in Table \ref{S - Table A}. The mean, median and the most probable value of $\langle H_{\rm mag} / H_{\rm th} \rangle_{MC}$ are 23.68, 14.13 and 4.78 respectively. The mean, median and most probable value of $\langle H_{\rm mag} / H_{\rm th} \rangle_{BG1}$ are 2.63 , 1.73 and 1.10 respectively while for $\langle H_{\rm mag} / H_{\rm th} \rangle_{BG2}$ they are are 2.84, 2.12 and 1.63 respectively. The mean value of each histogram is marked by a vertical line.}
              
         \label{Figure Hm_Hg}
   \end{figure}\\ 
\section{Comparing $H$ inside MCs and the background solar wind}
\label{S - H in MC and bg}
The \textit{in-situ} data from WIND spacecraft (\url{https://wind.nasa.gov/}) provides a detailed time profile of all the components of the plasma velocity ($\mathbf{v}$), proton number density ($n$) and all components of the magnetic field ($\mathbf{B}$)  in the spacecraft reference frame. We will use the terms ``plasma'' and ``magnetized fluid'' interchangeably in the rest of the paper; therefore the term ``plasma velocity'' can be taken to mean the quantity ${\mathbf v}$ introduced in \S~\ref{S - Bernoulli}. The data allow us to also compute the angle ($\theta$) between the plasma velocity $\mathbf{v}$ and the magnetic field $\mathbf{B}$. The data also provide the time profile of the plasma thermal pressure ($P_{\rm th}$) which includes contributions from protons and electrons (\url{https://omniweb.gsfc.nasa.gov/ftpbrowser/bow_derivation.html}). Assuming the electron number density to be equal to the proton number density (so that $\rho = n (m_{p} + m_{e})$ where $m_{p}$ and $m_{e}$ are the proton and electron masses respectively) and adopting two values for the polytropic index ($\gamma = 5/3$ and 1.2), the data allow us to calculate $H$ (Eq~\ref{eq: totH}) along the line of spacecraft intercept. 

\subsection{$\langle H \rangle_{MC}$ vs $\langle H \rangle_{BG}$}
\label{Sub - hmc_vs_hbg}
We compare the average value of $H$ inside MCs with that in the background solar wind using the quantity

\begin{equation}
H_{\rm R} \equiv \frac{\langle H \rangle_{MC}}{\langle H \rangle_{BG}}
\label{eq:HR}
\end{equation}
\begin{figure}
	\centering
	\includegraphics[width=\hsize]{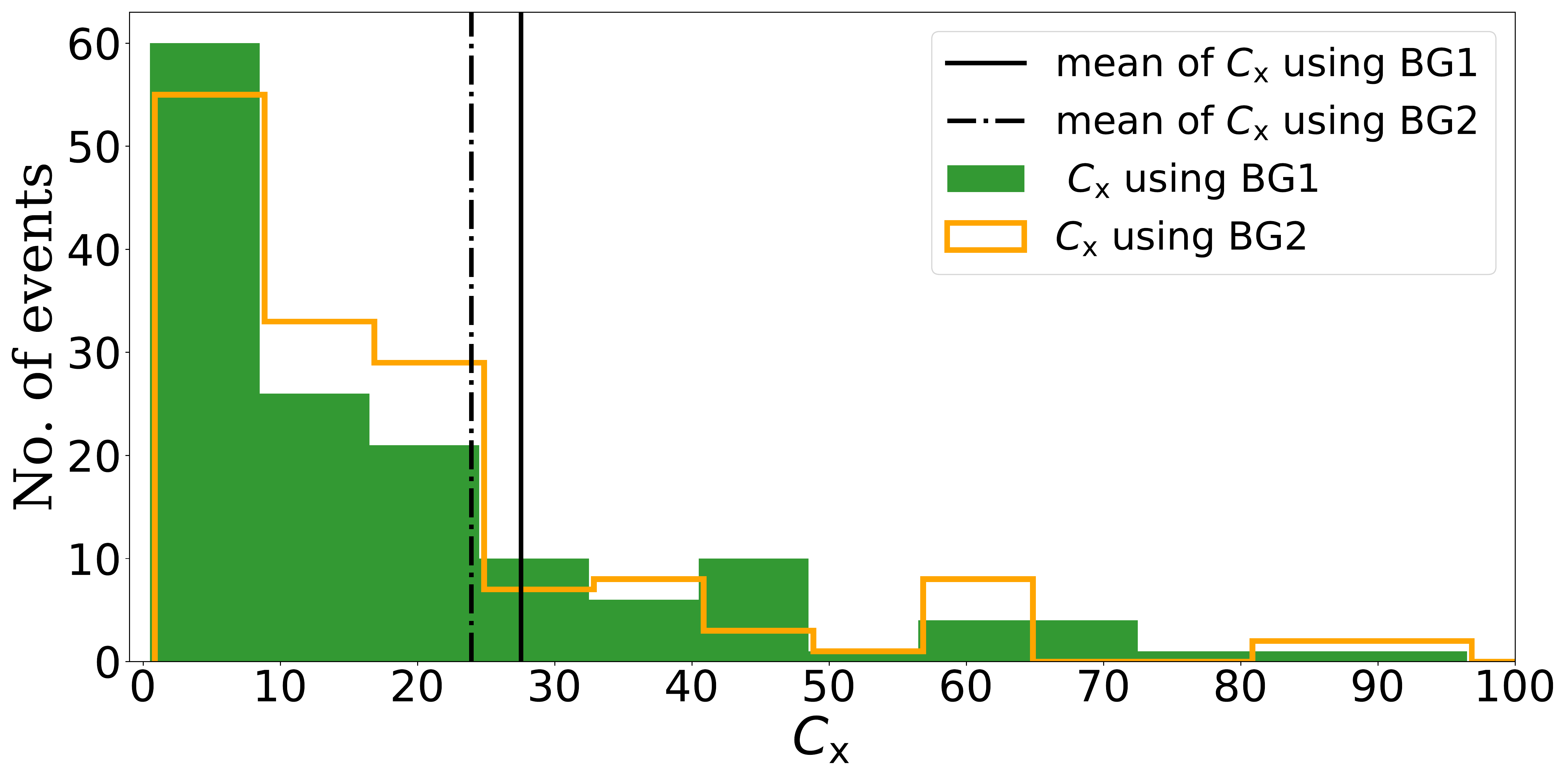}
	\caption{Histograms of $C_{\rm x}$ (Equation \ref{eq: Cx}) with $\gamma = 5/3$ and using the two backgrounds, BG1 and BG2. The mean, median and the most probable value of $C_{\rm x}$ using BG1 are 27.53, 11.09 and 2.48 respectively while for BG2, they are 23.91, 14.02 and 4.36 respectively. The mean value for each histogram is marked by a vertical line. The maximum value shown on the x axis is limited to 100 for zooming in on the histogram peaks.}
	
	\label{Figure Cx}
\end{figure}
\begin{figure}[h]
	\centering
	\includegraphics[width=\hsize]{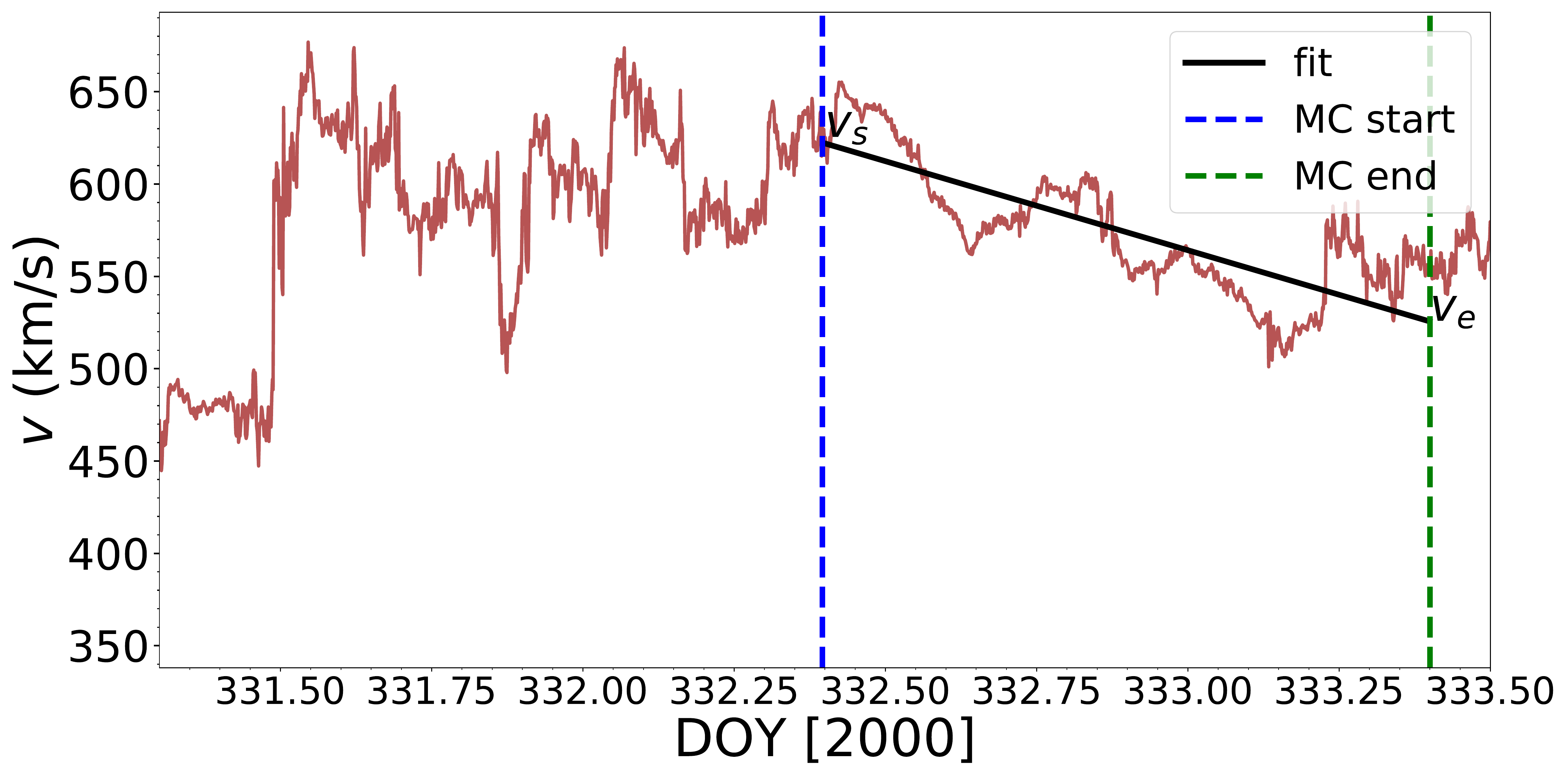}
	\caption{Velocity ($v$) profile as a function of time for event 54 in our dataset (Table~\ref{S - Table A}). The start time ($t_{\rm s}$) and end time ($t_{\rm e}$) of the MC are marked by blue and green dashed lines respectively. The black solid line shows a linear fit to $v$. $v_{\rm s}$ and $v_{\rm e}$ denote the value of the fit at $t_{\rm s}$ and $t_{\rm e}$ respectively. We compute the MC expansion speed $v_{\rm exp}$ (Equation~\ref{eq: vexp}) using $v_{\rm s}$ and $v_{\rm e}$.}
	
	\label{Figure vexp}
\end{figure}    
where $\langle \, \rangle_{MC}$ and $\langle \, \rangle_{BG}$ represent averages inside the MC and the background respectively. We compute $H_R$ using two different backgrounds (BG1 and BG2) which are described in \S~\ref{S - data}. We first describe the results obtained using $\gamma = 5/3$. The histograms of $H_{\rm R}$ for all the events in our sample is shown in the top panel of Figure~\ref{Figure H}. The mean, median and the most probable value of $H_{\rm R}$ using BG1 are 1.26, 1.1 and 0.97 respectively. Using BG2 they are 1.36, 1.19 and 1.12 respectively. 
The bottom panel of Figure~\ref{Figure H} shows a histogram of the ratio $\langle H_{\rm k}/H \rangle$ inside the MC and in the two  backgrounds, BG1 and BG2. The mean, median and most probable value of $\langle H_{\rm k}/H \rangle_{MC}$ are 0.90, 0.92 and 0.94 respectively. The mean, median and the most probable values of $\langle H_{\rm k}/H \rangle_{BG}$ are not very different; they are 0.98, 0.99 and 0.995 respectively for BG1 and 0.99, 0.99 and 0.996 for BG2. We note that though BG1 and BG2 are selected based on different criteria, the results are quite similar. 
Using $\gamma = 1.2$, the mean, median and most probable values of $H_{\rm R}$ are 1.27, 1.09 and 0.93 respectively (using BG1) and 1.36, 1.19 and 1.14 respectively (using BG2). Using $\gamma = 1.2$ we find that the mean, median and most probable value of $\langle H_{\rm k}/H \rangle_{MC}$ are 0.89, 0.91 and 0.92 respectively. The mean, median and most probable value of $\langle H_{\rm k}/H \rangle_{BG}$ for both the backgrounds are 0.98, 0.99 and 0.992 respectively. 
The main conclusions at this point are i) the total specific energy ($H$) inside the MC is approximately the same as that in the background, ii) $H_{k} \approx H$, both inside the MC and in the background. iii) the choice of the background and polytropic index does not affect these broad conclusions.

\begin{figure}[h]   
	
	\centering
	
	\includegraphics[width=\hsize]{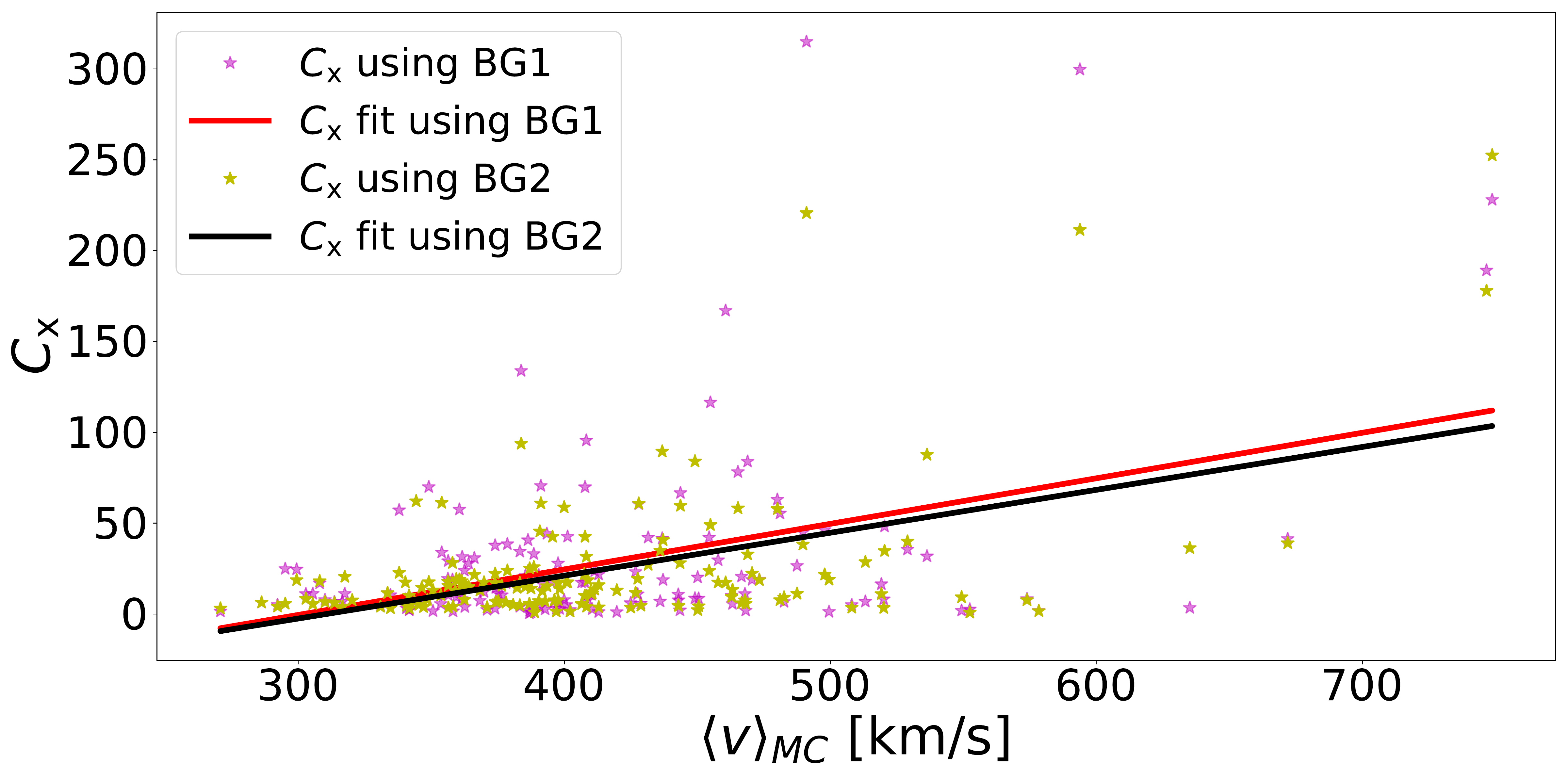}
	\includegraphics[width=\hsize]{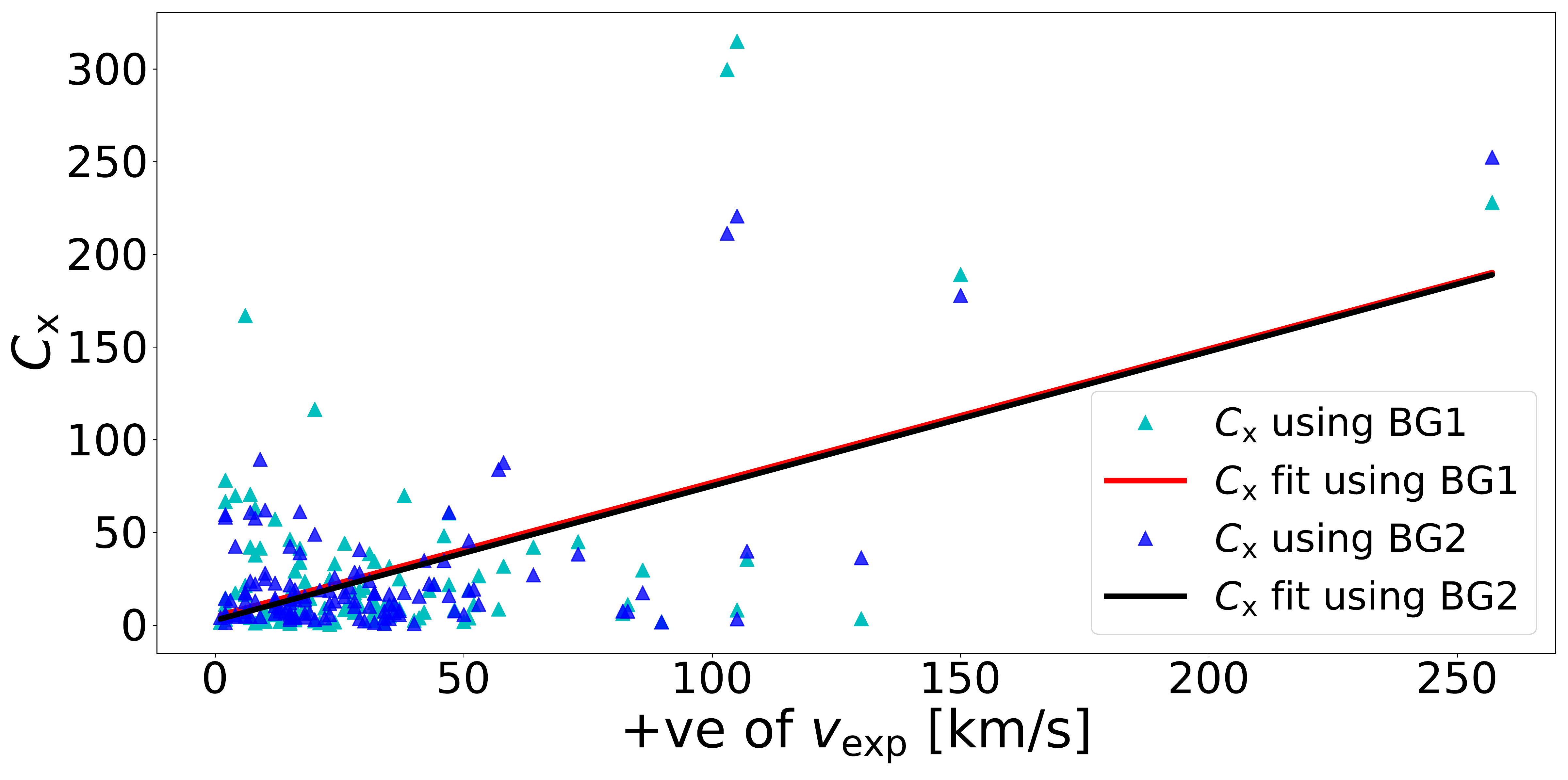}
	\includegraphics[width=\hsize]{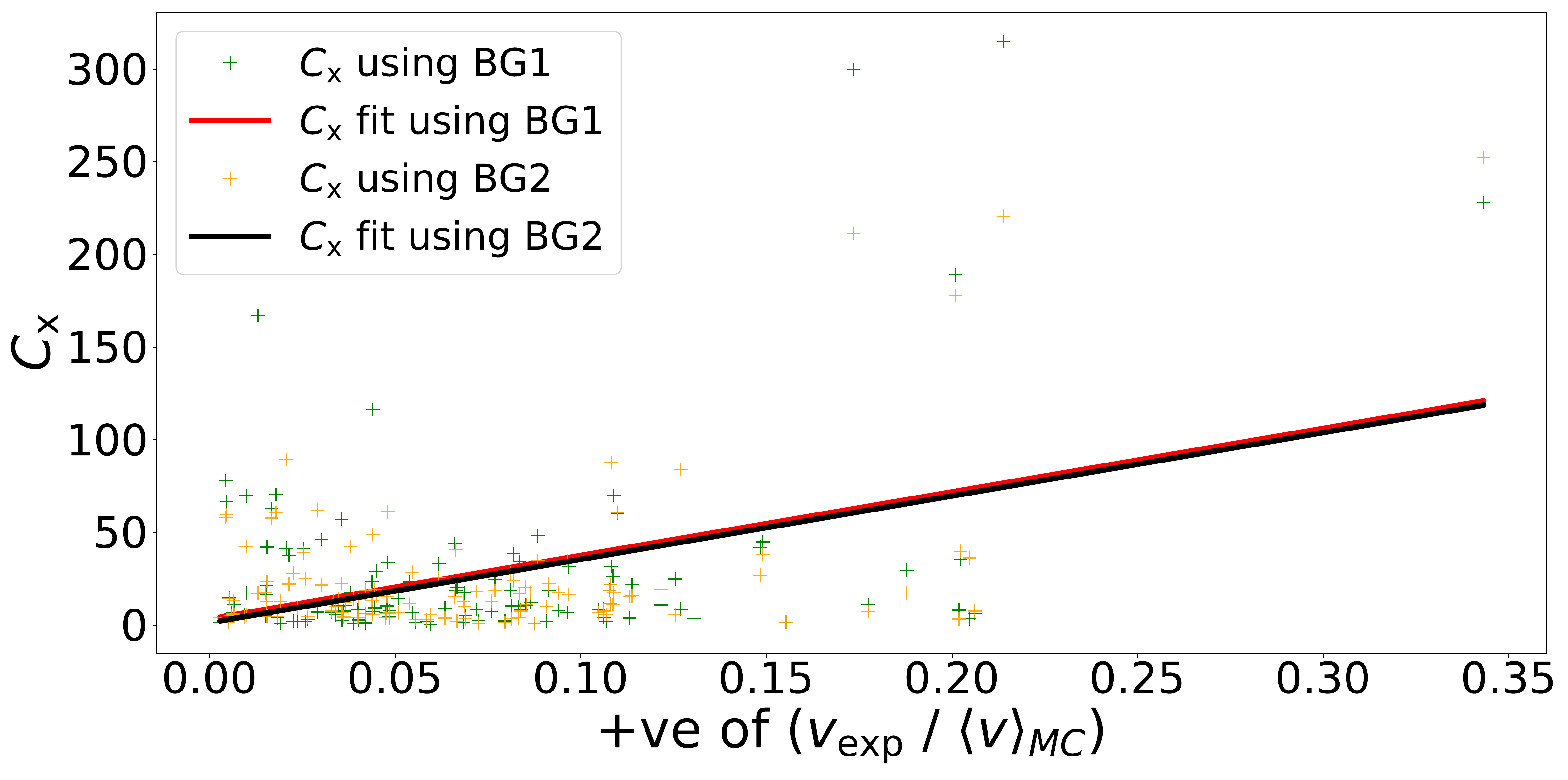}
	\caption{
			Scatterplots of $C_{\rm x}$ with MC propagation and expansion speeds for all the events listed in Table~\ref{S - Table A}. The top panel is the scatterplot between $C_{\rm x}$ (estimated using $\gamma = 5/3$) (Equation~\ref{eq: Cx}) and the MC propagation speed ($\langle v \rangle_{MC}$). The correlation coefficient between $C_x$ and $\langle v \rangle_{MC}$ are $r = 0.43$ with $p = 3.5\times 10^{-8}$ (for BG1) and $r = 0.51$ with $p = 1.8\times 10^{-11}$ (for BG2). The equation of the fitted lines corresponding to BG1 and BG2 are  $y = 0.25x - 75.65$ and $y = 0.24x - 73.28$ respectively. The middle panel is the scatterplot between $C_x$ (computed using $\gamma = 5/3$) and the MC expansion speed ($ v_{\rm exp}$), only for events with $v_{\rm exp} > 0$. The correlation coefficients are $r = 0.50$ with $p = 5.73\times 10^{-9}$ for BG1 and $r = 0.63$ with $p = 1.9\times 10^{-14}$ for BG2. The equation of fitted lines are $y = 0.72x + 4.82$ and $y = 0.73x + 2.73$ corresponding to BG1 and BG2 respectively. The bottom panel is the scatterplot between $C_x$ (calculated using $\gamma = 5/3$) and $v_{\rm exp}/\langle v \rangle_{MC}$ (only for events with $v_{\rm exp} > 0$). The correlation coefficients are $r = 0.40$ with $p = 7.8 \times 10^{-6}$(for BG1) and $r = 0.50$ with $p = 1.2\times 10^{-8}$(for BG2). The equation of the fitted lines are $y = 342.6x + 3.36$ (for BG1) and $y = 341.71x + 1.46$ (for BG2). The small p-values imply a high statistical confidence in computing the $r$ values in all the three cases.}
	\label{Figure Cx_speed_variation}
\end{figure}

\subsection{Comparing the thermal+magnetic specific energy inside MCs and the background}
\label{Sub - total specific energy}

Having shown that the contribution from the kinetic energy term ($H_{\rm k}$) dominates the specific energy both in the background solar wind and inside the MC, we now compare the thermal+magnetic contributions to the specific energy in the MC and the background using the metric
\begin{equation}
	C_{\rm x} \equiv \frac{{\huge \langle} \epsilon {\huge \rangle}_{MC}}{\langle \epsilon \rangle_{BG}} \, , \, \, {\rm where}\, \, \epsilon \equiv [H_{\rm th}^2\cos^2\theta + (H_{\rm th} + H_{\rm mag})^2\sin^2\theta]^{1/2}
	\label{eq: Cx}
\end{equation}
The values of $C_{\rm x}$ are listed in Table~\ref{S - Table B}. As mentioned earlier, $\langle \, \rangle_{MC}$ and $\langle \, \rangle_{BG}$ represent averages inside the MC and the background solar wind respectively. The quantity $\epsilon$ is the thermal + magnetic contribution to the specific energy (often called the enthalpy). The thermal contribution to the specific energy in an unmagnetized fluid is well known to be $\gamma P_{\rm th} / (\gamma -1)$, and the expression for $\epsilon$ (used in Eq~\ref{eq: Cx}) is written following the reasoning in Eq~\ref{eq: totH}. The histograms of $C_{\rm x}$ in Figure~\ref{Figure Cx} have a mean, median and most probable value of 27.53, 11.09 and 2.48 respectively (using BG1) and 23.91, 14.02 and 4.36 respectively (using BG2). The mean values (which are relatively high in comparison to the median and most probable value) are biased by $\approx 40 \%$ of events for both the backgrounds. Our results show that the thermal + magnetic specific energy inside the MC is generally higher than that of the background. If we use $\gamma = 1.2$ instead of 5/3, we find the mean, median and the most probable values of $C_{\rm x}$ are 18.48, 8.30 and 4.70 respectively (using BG1) and 16.51, 9.89 and 3.32 respectively (using BG2). The statistical picture of $C_{\rm x}$ thus do not differ significantly on using $\gamma = 1.2$. With $\gamma = 5/3$, the mean, median and the most probable value of $\langle H_{\rm mag}/H_{\rm th} \rangle$ inside MCs are 23.68, 14.13 and 4.78 respectively, while it is only 2.63, 1.73 and 1.10 respectively inside BG1 and 2.84, 2.12 and 1.63 inside BG2 (Figure~\ref{Figure Hm_Hg}). With $\gamma = 1.2$, $\langle H_{\rm mag}/H_{\rm th} \rangle$ is smaller (in comparison to the values with $\gamma = 5/3$) both inside the MC and in the backgrounds. The mean, median and the most probable value of $\langle H_{\rm mag}/H_{\rm th} \rangle_{MC}$ are 9.87, 5.89 and 2.43 respectively, and it is 1.10, 0.72 and 0.46 respectively (for BG1) and 1.19, 0.88 and 0.59 respectively (for BG2). The enhancement of $C_{\rm x}$ inside MCs is thus primarily due to the magnetic fields.  
\subsection{Comparing the thermal+magnetic pressure inside MCs and the background}
\label{Sub - total pressure}
\begin{figure}[h]
   \centering
   \includegraphics[width=\hsize]{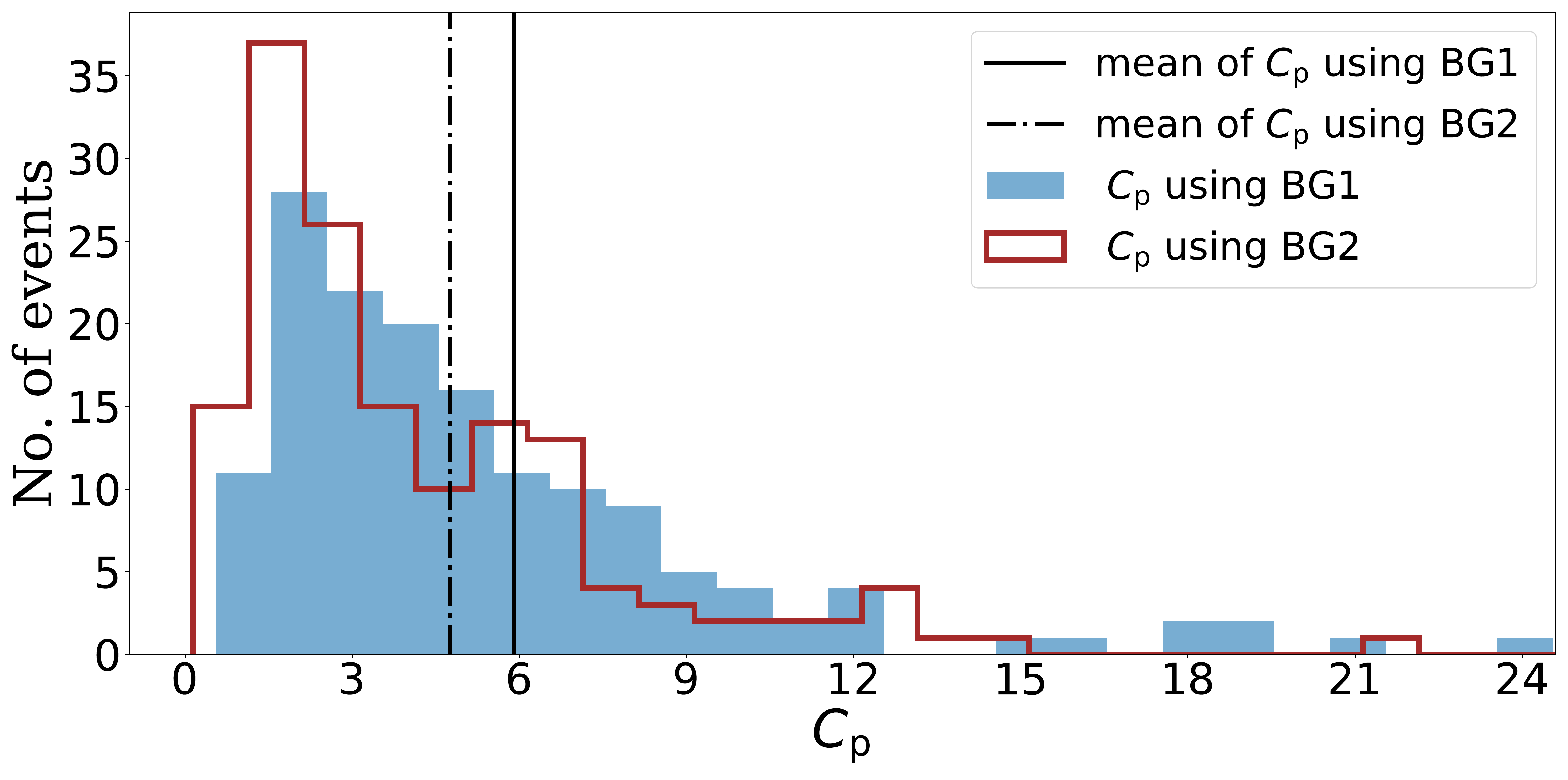}
      \caption{Histograms of $C_{\rm p}$ (Equation \ref{eq: Cp}) using two different backgrounds, BG1 and BG2. The mean, median and the most probable value of $C_{\rm p}$ with BG1 are 5.90, 4.29 and 2.39 respectively and the mean, median and the most probable value of $C_{\rm p}$ with BG2 are 4.76, 3.02 and 1.64 respectively. The mean value for each histogram is marked by a vertical line. The maximum value shown on the x axis is limited to 24.5 in order to zoom in on the histogram peaks.}
              
         \label{Figure Cp}
   \end{figure}

\begin{figure} [h!]  
	\centering
	
	\includegraphics[width=\hsize]{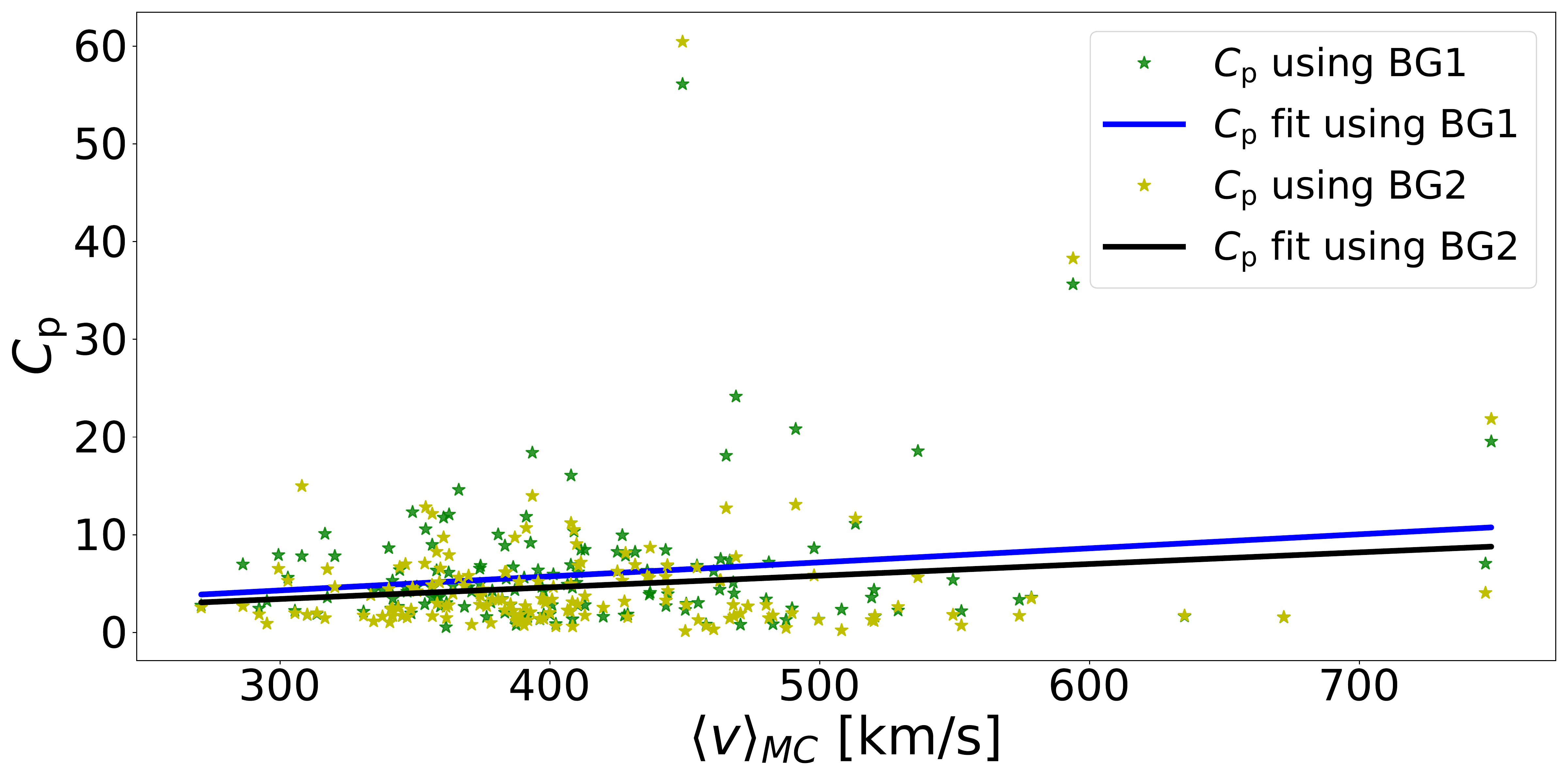}
	\includegraphics[width=\hsize]{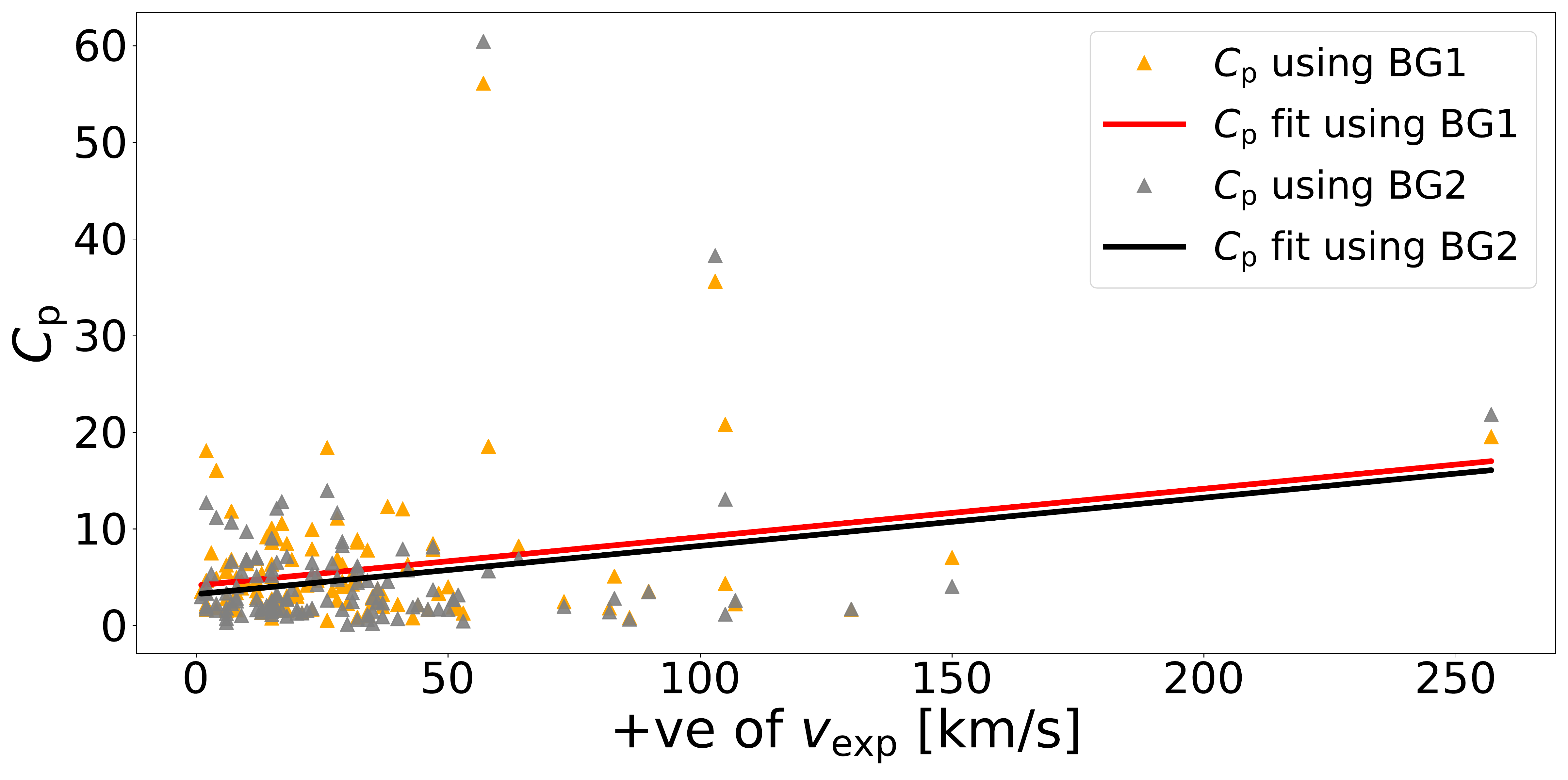}
	\includegraphics[width=\hsize]{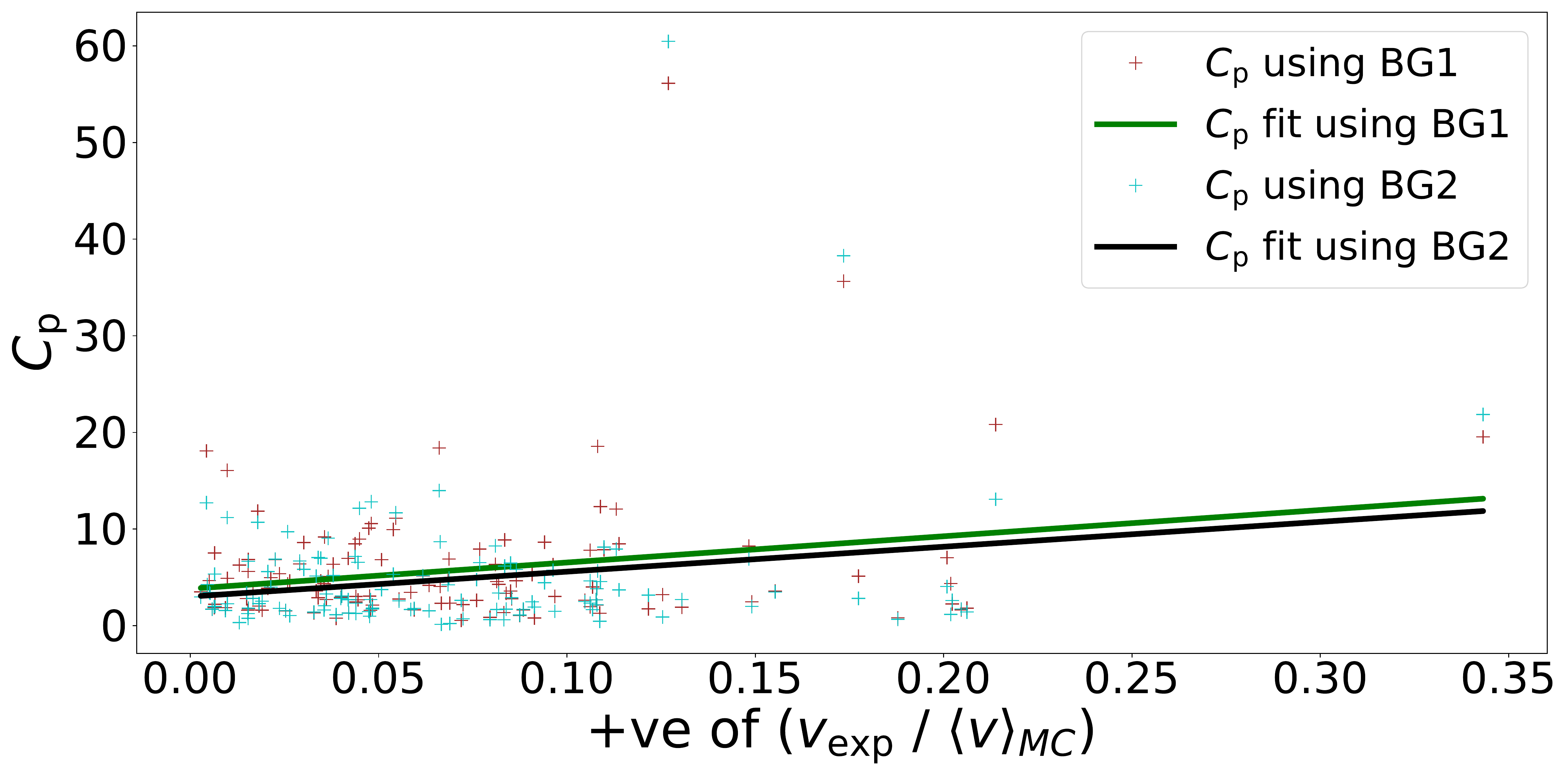}
	\caption{
			Scatterplots of $C_{\rm p}$ with MC propagation and expansion speeds for all the events listed in Table~\ref{S - Table A}. The top panel is the scatterplot between $C_{\rm p}$ (Equation~\ref{eq: Cp} and the MC propagation speed ($\langle v \rangle_{MC}$). The correlation coefficient between $C_{\rm p}$ and $\langle v \rangle_{MC}$ is $r = 0.18$ with $p = 2\times 10^{-2}$ (for BG1) and $r = 0.15$ with $p = 5\times 10^{-2}$ (for BG2). The equation of the fitted lines corresponding to BG1 and BG2 are  $y = 0.01x - 0.004$ and $y = 0.01x - 0.1$ respectively. The middle panel is the scatterplot between $C_{\rm p}$ and the MC expansion speed ($ v_{\rm exp}$), only for events with $v_{\rm exp} > 0$. The correlation coefficients are $r = 0.25$ with $p = 2\times 10^{-3}$ for BG1 and $r = 0.25$ with $p = 5\times 10^{-3}$ for BG2. The equation of the fitted lines are $y = 0.05x + 4.5$ and $y = 0.05x + 3.26$ for BG1 and BG2 respectively. The bottom panel is the scatterplot between $C_{\rm p}$ and $ v_{\rm exp}/\langle v \rangle_{MC}$, only for events with $v_{\rm exp} > 0$. The correlation coefficients are $r = 0.23$ with $p = 6 \times 10^{-3}$ (for BG1) and $r = 0.22$ with $p = 10^{-3}$ (for BG2). The equation of the fitted lines are $y = 25.3x + 4.30$ (for BG1) and $y = 25.78x + 3.02$ (for BG2). The small p-values indicate a sufficiently high statistical confidence in estimating the $r$ values for all the three cases.}
	\label{Figure Cp_speed_variation}
\end{figure}        
 There are several studies concerning the difference between the thermal+magnetic pressure $P_{\rm th} + P_{\rm mag}$ inside CMEs and the ambient solar wind \citep{1981BurlagaJGR, 2000MoldwinGeoRL, 2005JianESASP, 2009DDA&A, 2015GopalJGRA}. It is often speculated that the reason ICMEs expand internally is because they are overpressured with respect to the ambient solar wind \citep{2019ScoliniA&A}. \citet{2015GopalJGRA} and \citet{2021Mishra} examined if ICMEs were more overpressured during relatively weaker solar cycles, leading to more halo CMEs during such cycles. 
We therefore  compute the following coefficient for the events in our sample using BG1 and BG2:
\begin{equation}
C_{\rm p} \equiv \frac{\langle P_{\rm mag} + P_{\rm th} \rangle_{MC}}{\langle P_{\rm mag} + P_{\rm th} \rangle_{BG}} 
\label{eq: Cp}
\end{equation}
where $P_{\rm mag}$ ($={\mathbf{B}}^{2}/8 \pi$) is the magnetic pressure and $P_{\rm th}$ is the thermal pressure of the plasma. The thermal pressure $P_{\rm th}$ includes contributions from both protons and electrons and $\langle \, \rangle_{MC}$ and $\langle \, \rangle_{BG}$ have their usual meanings. The $C_{\rm p}$ values are listed in Table~\ref{S - Table B} and histograms for $C_{\rm p}$ are shown in Figure \ref{Figure Cp}. The mean, median and the most probable value for $C_{\rm p}$ using BG1 are 5.90, 4.29 and 2.39 respectively and using BG2, they are 4.76, 3.02 and 1.64 respectively. By comparison, \citet{2015GopalJGRA} find that the total pressure ratio between the MCs and the ambient solar wind is $\approx 3$ for their set of near-Earth CMEs. The results thus suggest that the average magnetic+thermal pressure inside near-Earth MCs is appreciably higher than that of the solar wind background. The polytropic index $\gamma$ has no bearing on $C_{\rm p}$. 
\subsection{Is the excess thermal+magnetic specific energy and pressure inside MCs correlated with near-Earth expansion and propagation speeds?}
\label{Sub - cxcpwithvexp}
Just as several studies ask if the excess pressure inside ICMEs leads to their expansion \citep[see e.g.,][]{2006vonSteigerSSRv, 2019ScoliniA&A,2009DDA&A,2022Verbeke}, it is natural to ask if the enhanced thermal+magnetic specific energy inside MCs result in their expansion.
We compute the MC expansion speed \citep{18NCSo}
\begin{equation}
v_{\rm exp} = \frac{1}{2}(v_{\rm s} - v_{\rm e})
\label{eq: vexp}
\end{equation}
for each MC in our sample. The quantities $v_{\rm s}$ and $v_{\rm e}$ are the speeds at the start and at the end of the MC respectively. Figure~\ref{Figure vexp} shows an example of a linear fit that is used to compute $v_{s}$, $v_{e}$ and consequently $v_{\rm exp}$. We note that $v_{\rm exp}$ for $\approx 20\%$ of MCs in our sample is negative; i.e., they contract, rather than expand. Figure~\ref{Figure Cx_speed_variation} shows scatterplots between $C_{\rm x}$ (estimated using both BG1 and BG2) and the near-Earth MC propagation speed $\langle v \rangle_{MC}$ (panel a) and $v_{\rm exp}$ (panel b) for expanding MCs (i.e.; those for which $v_{\rm exp} > 0$). 
The linear correlation coefficient ($r$) between $C_{\rm x}$ and the MC propagation speed ($\langle v \rangle_{MC}$) is 0.43 with a p-value of $3.5 \times 10^{-8}$ with BG1 and $r = 0.51$ with $p = 1.8 \times 10^{-11}$ with BG2. The linear correlation coefficient ($r$) between $C_x$ and $v_{\rm exp}$ for the expanding events is 0.50 and the corresponding p-value is $5.73 \times 10^{-9}$ (using BG1) and $r = 0.63$ with $p = 1.9 \times 10^{-14}$ (using BG2). The correlation coefficient between $C_{\rm x}$ and $ v_{exp}/\langle v \rangle$ (considering only events with positive $v_{\rm exp}$) $r = 0.40$ with a $p$-value $7.8 \times 10^{-6}$ for BG1 and $r = 0.50$ with $p = 1.2 \times 10^{-8}$ for BG2. 
Our findings suggest that $C_{\rm x}$ is only moderately correlated with the near-Earth MC propagation and expansion speed. The low values for $p$ in all the cases indicate a high statistical significance for these results. 

We next study if/how the overpressure parameter $C_{\rm p}$ is correlated with the near-Earth MC expansion and propagation speeds. Panel (a) of Figure~\ref{Figure Cp_speed_variation} shows the scatterplot between $C_{\rm p}$ (using both BG1 and BG2) and the MC propagation speed. The correlation is low ($r = 0.18$, p-value = $2 \times 10^{-2}$ using BG1 and $r = 0.15$, $p = 5 \times 10^{-2}$ using BG2). The correlation between $C_{\rm p}$ and the MC expansion speed of the expanding MCs ($v_{\rm exp} > 0$) is also low (panel b of Figure~\ref{Figure Cp_speed_variation}, $r = 0.25$, p-value = $2 \times 10^{-3}$ using BG1 and $r = 0.25$ with $p = 5\times 10^{-3}$). Finally, we note that the correlation coefficient between $C_{\rm p}$ and $v_{\rm exp}/\langle v \rangle_{MC}$ for the expanding MCs (panel c of Figure~\ref{Figure Cp_speed_variation}) are $r = 0.23$ and the $p$-value is $6\times 10^{-3}$ for BG1 and $r = 0.22$ with $p = 10^{-3}$ for BG2. 
Evidently, $C_{\rm p}$ is rather poorly correlated with MC expansion and propagation speeds.
This might be because we are using speeds measured at the position of the WIND observation, whereas most of the CME expansion probably occurs closer to the Sun \citep{2022Verbeke}. 


\section{Does the excess thermal+magnetic specific energy in MCs make them appear ``rigid''?}
\label{S - rigidity}

\begin{figure}
   \centering
   \includegraphics[width=\hsize]{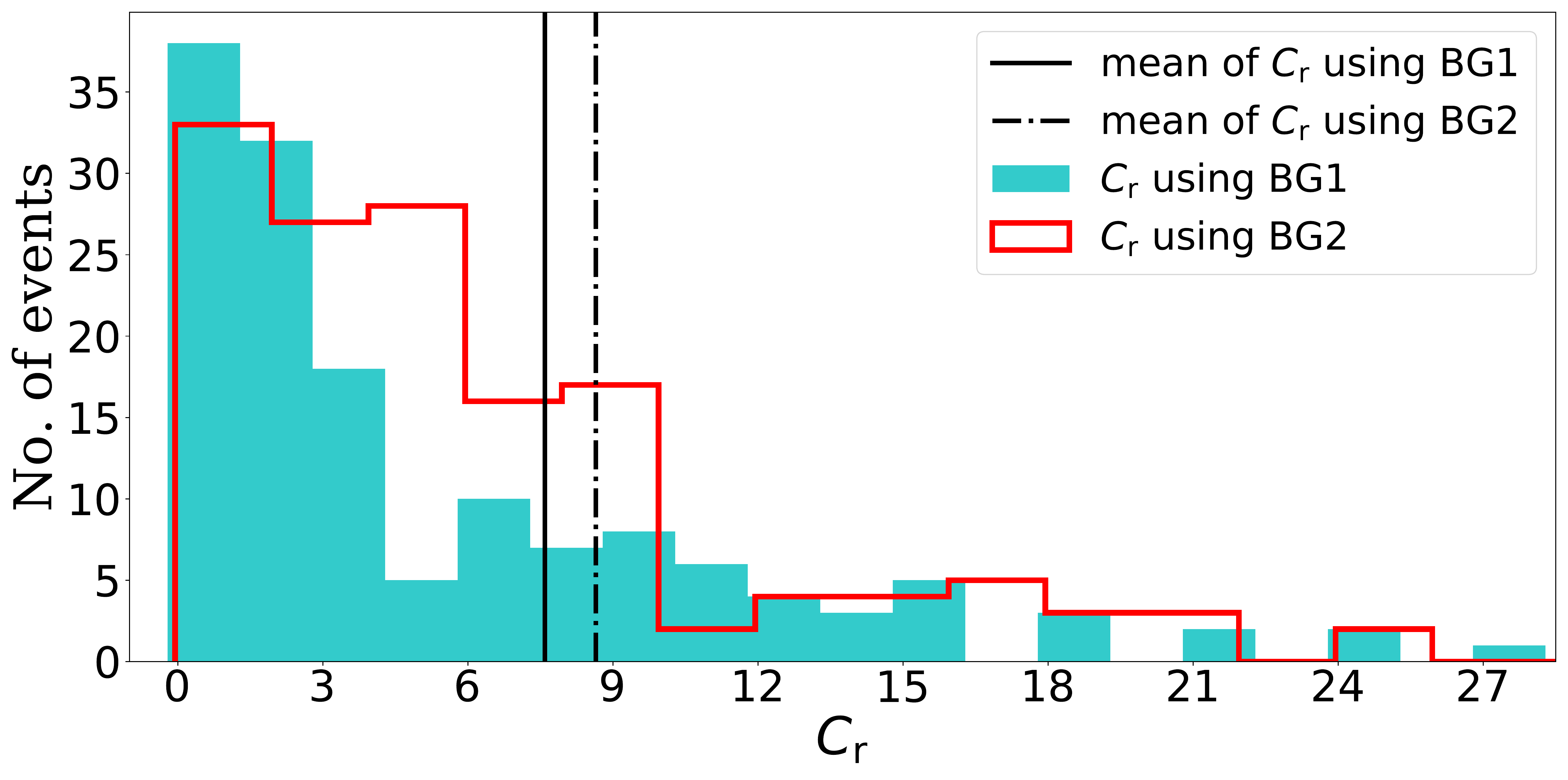}
      \caption{Histograms of $C_{\rm r}$ (Equation \ref{eq: Cr}) with $\gamma = 5/3$ using both the backgrounds (BG1 and BG2). The mean, median and the most probable value of $C_{\rm r}$ are 7.61, 3.33 and 1.16 respectively for BG1 and 8.65, 4.65 and 2.25 respectively for BG2. The mean value of each histogram is marked by a vertical line. The maximum value shown on the x axis is limited to 28 for zooming in on the histogram peaks.}
              
         \label{Figure Cr}
   \end{figure}

The main results we have obtained until now are that i) the total specific energy inside the MC $\approx$ that in the background, and ii) the sum of the thermal and magnetic (specific) energies is higher inside the MC as compared to the background. By way of trying to understand point ii) better, we define the following coefficient: 
\begin{equation}
C_{\rm r} = \frac{\langle \epsilon \rangle_{MC} - \langle \epsilon \rangle_{BG}}{\langle (1/2) u^2 \rangle_{BG}} \, ,
\label{eq: Cr}
\end{equation}
where $\epsilon$ is defined in Eq~\ref{eq: Cx}, $\mathbf{u} \equiv \mathbf{v} - \langle \mathbf{v} \rangle_{MC}$ is the solar wind velocity in the frame of the MC and the other notations carry their usual meanings. 
The quantity $\langle (1/2) u^2 \rangle_{BG}$ denotes the specific kinetic energy of the oncoming solar wind incident on the MC as discerned by an observer moving with the average MC speed. The quantity $C_{\rm r}$ (Equation~\ref{eq: Cr}) thus compares the excess thermal+magnetic specific energy in the MC (relative to the background) with the specific kinetic energy of the solar wind impinging on it. If $C_{\rm r} \gtrsim 1$, it means that the excess thermal + magnetic specific energy in the MC is greater than the specific kinetic energy of the oncoming solar wind - suggesting that the MC can resist deformation by the solar wind, somewhat like a rigid body. A way to understand $C_{\rm r}$ is as follows: imagine an inflated balloon placed in a stream of cold air. There are no magnetic fields, and so $H_{\rm mag} = 0$. Since the air stream is cold, $\langle H_{\rm th} \rangle_{\rm BG}$ is zero and the only term in the numerator of Eq~\ref{eq: Cr} is $\langle H_{\rm th} \rangle_{\rm MC}$. The metric $C_{\rm r}$ thus compares the thermal specific energy due of the gas inside the balloon with the specific kinetic energy of the cold air stream incident on it. If $C_{\rm r} \gtrsim 1$, the balloon can resist deformation due to the incident air stream relatively better (i.e., it behaves more like a well-inflated soccer ball) and vice-versa.

The $C_{\rm r}$ values for all the events in our dataset are listed in Table~\ref{S - Table B}. The histogram of $C_{\rm r}$ for all the events using $\gamma = 5/3$ and both the backgrounds (BG1 and BG2) is shown in Figure~\ref{Figure Cr}. With BG1, we find that the mean, median and most probable values of $C_{\rm r}$ histogram are 7.61, 3.33 and 1.16 respectively, and 81\% of the events have $C_{\rm r} \geq 1$. Using BG2, the mean, median and most probable values are 8.65, 4.66 and 2.25 respectively and 89.5\% of the events have $C_{\rm r} \geq 1$.  
If we assume $\gamma = 1.2$ (instead of 5/3), the mean, median and most probable values of $C_{\rm r}$ are 8.09, 3.55 and 1.12 respectively with BG1, and 9.50, 5.29 and 2.29 respectively with BG2. The choice of $\gamma$ does not affect our results substantially.

Furthermore, the magnetic energy is typically higher than the thermal energy inside MCs, (Figure~\ref{Figure Hm_Hg}) suggesting that CME magnetic fields are primarily responsible for $C_{\rm r} \geq 1$. 
 Our findings are in keeping with those of \cite{2011Lynnyk} who state ``If the magnetic ﬁeld inside the ICME/MC is much stronger than that in the ambient solar wind, the ICME/MC cross-section is closer to circular." MCs with more circular cross sections are probably ones which are relatively more rigid; i.e., ones which probably have $C_{\rm r} \geq 1$, and have resisted deformation by the solar wind to more elongated shapes.

An issue related to the rigidity of MCs is the widely used CME aerodynamic drag law \citep{2004CargillSoPh, 2009VrsIAUS, 2013VrsSoPh} 
\begin{equation}
F_{\rm D} = C_{\rm D} A \rho \mid u \mid u 
\label{eq: drag}
\end{equation}
which is a hydrodynamic drag law applicable to high Reynolds number flows past {\em solid/rigid bodies} \citep{1987Landau}. In Equation~\ref{eq: drag}, $F_{\rm D}$ is the drag force, $u$ is the velocity of the background fluid as viewed by an observer comoving with the body, $A$ is the cross-sectional area subtended by the body to the flow, $\rho$ is the mass density of the background fluid and $C_{\rm D}$ is a dimensionless proportionality constant. By contrast, the high Reynolds number drag law for flows past deformable bubbles looks like $F_{\rm D} \propto u$ \citep{1987Landau,1959MooreJFM,1963MooreJFM,1988KangPhFl}. Although the $F_{\rm D} \propto u$ resembles the Stokes law, which is applicable for laminar/low Reynolds number flows past solid bodies, it is in fact a high Reynolds number law for flows past deformable bubbles. Bubbles are distinct from solid bodies in that the total velocity does not vanish on their surface, as it does for solid bodies. We note that some studies of CME dynamics do adopt $F_{\rm D} \propto u$ \citep{2002VrsnakJGRA,2010MaloneyApJ}. Although the form of the widely adopted CME aerodynamic drag law arises from an amalgamation of several MHD effects \citep{2022LinJGRA}, the simplest interpretation of Eq~\ref{eq: drag} is still that of a high Reynolds number, solid/rigid body law. Our finding $C_{\rm r} \gtrsim 1$ (for over 89 \% of the MCs we study) might be a possible justification for the solid body premise, even for CMEs, which are obviously far from solid bodies in the usual sense of the word.
 


\section{Conclusions and discussion}
\label{S - conclusions}
CME evolution through the heliosphere is thought to be strongly influenced by the difference in the (thermal + magnetic) pressures inside the CME and outside, in the ambient solar wind. In this paper we compare the average value of the total specific energy $H$ (Eq~\ref{eq: totH}) inside MCs and the background solar wind using the near-Earth {\em in-situ} data from the WIND spacecraft for a set of 152 well observed MCs. The quantity $H$ contains contributions from the kinetic energy due to the bulk flow ($H_{\rm k}$), the thermal energy ($H_{\rm th}$) and the magnetic energy ($H_{\rm mag}$). We also compare the thermal+magnetic pressure inside MCs with the background solar wind. We use two different ambient solar wind backgrounds for our comparisons and also use two different values for the polytropic index $\gamma$. Our main conclusions are as follows: 

\begin{enumerate}
\item
The average value of $H$ inside MCs, $\langle H \rangle_{\rm MC}$ $\approx$ the average value in the ambient solar wind $\langle H \rangle_{\rm BG}$ (\S~\ref{Sub - hmc_vs_hbg}, Figure~\ref{Figure H}). The bulk flow kinetic energy contribution $H_{\rm k} \approx H$, both inside the MCs and in the ambient solar wind. This is the primary reason for $\langle H \rangle_{\rm MC} \approx \langle H \rangle_{\rm BG}$.
\item
The average thermal+magnetic specific energy inside near-Earth MCs substantially exceeds that in the background solar wind (\S~\ref{Sub - total specific energy}). Similarly, the average $P_{\rm th} + P_{\rm mag}$ inside near-Earth MCs is greater than that in the solar wind background (\S~\ref{Sub - total pressure}). These conclusions are broadly consistent with the findings of \citet{2015GopalJGRA} and \citet{2021Mishra}.
\item
We find that the excess thermal+magnetic energy inside MCs is moderately correlated with MC near-Earth propagation and expansion speeds, while the correlation is rather poor for the excess thermal+magnetic pressure (\S~\ref{Sub - cxcpwithvexp}). 
Summarizing, neither the excess enthalpy nor total pressure seems to be well correlated with the near-Earth MC propagation and expansion speeds. This might be because most of the expansion occurs closer to the Sun \citep{1999OdstrcilJGR,2014GopalGeoRL,2022Verbeke} and/or because magnetic field rearrangement is the primary reason for expansion \citep{1996KumarRustJGR}.

\item
We find that the excess thermal+magnetic specific energy inside MCs $\gtrsim$ the specific kinetic energy of the solar wind impinging on them for 81--89\% of the events we study (\S~\ref{S - rigidity}). This suggests how MCs might able to resist deformation by the solar wind, and might suggest a justification for the popular ``rigid body'' CME aerodynamic drag law (Equation~\ref{eq: drag}).
\end{enumerate}

Our results are based only on {\em in-situ} data at the position of the WIND measurements (i.e., at 1 AU). It would be interesting to carry out similar calculations using {\em in-situ} data closer to the Sun from the PSP \citep{2020WeissEGUGA,2022NCApJ}. This would yield insights on the evolution of CMEs through the heliosphere.

\begin{acknowledgement}

DB acknowledges the PhD studentship from Indian Institute of Science Education and Research, Pune. We acknowledge a detailed review by an anonymous referee that helped us improve the manuscript.

\end{acknowledgement}

%

\begin{thebibliography}{}

\bibitem[\protect\citeauthoryear{Bhattacharjee et al.}{2022}]{2022Debesh} Bhattacharjee D., Subramanian P., Bothmer V., Nieves-Chinchilla T., Vourlidas A., 2022, SoPh, 297, 45. doi:10.1007/s11207-022-01982-x

\bibitem[Bothmer \& Schwenn (1998)]{1998BthAnGeo} Bothmer V., Schwenn R., 1998, AnGeo, 16, 1. doi:10.1007/s00585-997-0001-x

\bibitem[\protect\citeauthoryear{Boyd \& Sanderson}{2003}]{2003BoydSanderson} Boyd T.~J.~M., Sanderson J.~J., 2003, phpl.book, 544


\bibitem[Burlaga \emph{et al.}(1981)]{1981BurlagaJGR}Burlaga, L., Sittler, E., Mariani, F., and Schwenn, R.: 1981, {\it Journal of Geophysical Research} {\bf 86}, 6673. doi:10.1029/JA086iA08p06673.

\bibitem[Cargill et al.(1995)]{1995CargillGeoRL} Cargill, P.~J., Chen, J., Spicer, D.~S., et al.\ 1995, \grl, 22, 647. doi:10.1029/95GL00013  
  
\bibitem[Cargill, \& Schmidt(2002)]{2002CargillAnGeo} Cargill, P.~J., \& Schmidt, J.~M.\ 2002, Annales Geophysicae, 20, 879  
  
\bibitem[Cargill(2004)]{2004CargillSoPh} Cargill, P.~J.\ 2004, \solphys, 221, 135  
  
\bibitem[\protect\citeauthoryear{Chan{\'e} et al.}{2021}]{2021ChaneA&A} Chan{\'e} E., Schmieder B., Dasso S., Verbeke C., Grison B., D{\'e}moulin P., Poedts S., 2021, A\&A, 647, A149. doi:10.1051/0004-6361/202039867
  
\bibitem[\protect\citeauthoryear{Chen \& Garren}{1993}]{1993ChenGarren} Chen J., Garren D.~A., 1993, GeoRL, 20, 2319. doi:10.1029/93GL02426
  
  
\bibitem[Chen(1996)]{1996ChenJGR}Chen, J.: 1996, {\it Journal of Geophysical Research} {\bf 101}, 27499. doi:10.1029/96JA02644.
  
  
\bibitem[Chen(2011)]{2011ChenLRSP}Chen, P.F.: 2011, {\it Living Reviews in Solar Physics} {\bf 8}, 1. doi:10.12942/lrsp-2011-1.
  

  
\bibitem[Dakeyo \emph{et al.}(2022)]{2022Dakeyo}Dakeyo, Maksimovic, D{\'e}moulin, Halekas, and Stevens: 2022, {\it arXiv e-prints}, arXiv:2207.03898.
  
  
  \bibitem[D{\'e}moulin and Dasso(2009)]{2009DDA&A}D{\'e}moulin, P. and Dasso, S.: 2009, {\it Astronomy and Astrophysics} {\bf 498}, 551. doi:10.1051/0004-6361/200810971.
  
\bibitem[\protect\citeauthoryear{Davies et al.}{2021}]{2021DaviesA&A} Davies E.~E., M{\"o}stl C., Owens M.~J., Weiss A.~J., Amerstorfer T., Hinterreiter J., Bauer M., et al., 2021, A\&A, 656, A2. doi:10.1051/0004-6361/202040113
  

\bibitem[\protect\citeauthoryear{Dasso et al.}{2007}]{2007DassoSoPh} Dasso S., Nakwacki M.~S., D{\'e}moulin P., Mandrini C.~H., 2007, SoPh, 244, 115. doi:10.1007/s11207-007-9034-2

  
  \bibitem[Forsyth \emph{et al.}(2006)]{2006ForsythSSRv}Forsyth, R.J., Bothmer, V., Cid, C., Crooker, N.U., Horbury, T.S., Kecskemety, K., and, ...: 2006, {\it Space Science Reviews} {\bf 123}, 383. doi:10.1007/s11214-006-9022-0.


\bibitem[Gopalswamy \emph{et al.}(2014)]{2014GopalGeoRL}Gopalswamy, N., Akiyama, S., Yashiro, S., Xie, H., M{\"a}kel{\"a}, P., and Michalek, G.: 2014, {\it Geophys. Res. Lett.} {\bf 41}, 2673. doi:10.1002/2014GL059858.

  
  \bibitem[Gopalswamy \emph{et al.}(2015)]{2015GopalJGRA}Gopalswamy, N., Yashiro, S., Xie, H., Akiyama, S., and M{\"a}kel{\"a}, P.: 2015, {\it Journal of Geophysical Research (Space Physics)} {\bf 120}, 9221. doi:10.1002/2015JA021446.

\bibitem[Groth \emph{et al.}(2000)]{2000GrothJGR}Groth, C.P.T., De Zeeuw, D.L., Gombosi, T.I., and Powell, K.G.: 2000, {\it Journal of Geophysical Research} {\bf 105}, 25053. doi:10.1029/2000JA900093.


\bibitem[Guo \emph{et al.}(2011)]{2011GuoJGRA}Guo, J., Feng, X., Emery, B.A., Zhang, J., Xiang, C., Shen, F., and, ...: 2011, {\it Journal of Geophysical Research (Space Physics)} {\bf 116}, A05106. doi:10.1029/2011JA016490.


\bibitem[\protect\citeauthoryear{Hidalgo}{2003}]{2003HidalgoJGRA} Hidalgo M.~A., 2003, JGRA, 108, 1320. doi:10.1029/2002JA009818

   
   \bibitem[Jian \emph{et al.}(2005)]{2005JianESASP}Jian, L., Russell, C.T., Gosling, J.T., and Luhmann, J.G.: 2005, {\it Solar Wind 11/SOHO 16, Connecting Sun and Heliosphere} {\bf 592}, 731.

\bibitem[\protect\citeauthoryear{Jian et al.}{2008}]{2008JianAd} Jian L., Russell C.~T., Luhmann J.~G., Skoug R.~M., 2008, AdSpR, 41, 259. doi:10.1016/j.asr.2007.03.023


\bibitem[Kang and Leal(1988)]{1988KangPhFl}Kang, I.S. and Leal, L.G.: 1988, {\it Physics of Fluids} {\bf 31}, 233. doi:10.1063/1.866852.

\bibitem[Kassa Dagnew \emph{et al.}(2022)]{2022Dagnew}Kassa Dagnew, F., Gopalswamy, N., Belay Tessema, S., Akiyama, S., and Yashiro, S.: 2022, {\it arXiv e-prints}, arXiv:2208.03536.


\bibitem[Keppens \emph{et al.}(2020)]{2020Keppens}Keppens, R., Teunissen, J., Xia, C., and Porth, O.: 2020, {\it arXiv e-prints}, arXiv:2004.03275.


\bibitem[Kilpua \emph{et al.}(2013)]{2013KilpuaEGUGA}Kilpua, E., Isavnin, A., Vourlidas, A., Koskinen, H., and Rodriguez, L.: 2013, {\it EGU General Assembly Conference Abstracts}.

     
  \bibitem[Klein and Burlaga(1982)]{1982KleinBurlagaJGR}Klein, L.W. and Burlaga, L.F.: 1982, {\it Journal of Geophysical Research} {\bf 87}, 613. doi:10.1029/JA087iA02p00613.     

\bibitem[\protect\citeauthoryear{Klein et al.}{2012}]{2012KleinApJ} Klein K.~G., Howes G.~G., TenBarge J.~M., Bale S.~D., Chen C.~H.~K., Salem C.~S., 2012, ApJ, 755, 159. doi:10.1088/0004-637X/755/2/159
      
      
\bibitem[Kumar and Rust(1996)]{1996KumarRustJGR}Kumar, A. and Rust, D.M.: 1996, {\it Journal of Geophysical Research} {\bf 101}, 15667. doi:10.1029/96JA00544.
   
\bibitem[Kundu \& Cohen(2008)]{2008KunduCohen} Kundu, P.~K., \& Cohen, I.~M.\ 2008, Fluid Mechanics: Fourth Edition. Edited by Pijush K. Kundu and Ira M. Cohen with contributions by P. S. Ayyaswamy and H. H. Hu. ISBN 978-0-12-373735-9. Published by Academic Press
   
   
\bibitem[\protect\citeauthoryear{Kulsrud}{2005}]{2005Kulsrud} Kulsrud R.~M., 2005, \textit{Plasma Physics for Astrophysics}
   
   
\bibitem[Landau \& Lifshitz(1987)]{1987Landau} Landau, L.~D., \& Lifshitz, E.~M.\ 1987, Fluid Mechanics. Second Edition. 1987. Pergamon   
   
   \bibitem[\protect\citeauthoryear{Lepping et al.}{2003}]{2003LeppingSoPh} Lepping R.~P., Berdichevsky D.~B., Szabo A., Arqueros C., Lazarus A.~J., 2003, SoPh, 212, 425. doi:10.1023/A:1022938903870
   

\bibitem[Lin and Chen(2022)]{2022LinJGRA}Lin, C.-H. and Chen, J.: 2022, {\it Journal of Geophysical Research (Space Physics)} {\bf 127}, e28744. doi:10.1029/2020JA028744.

   
\bibitem[Linker \emph{et al.}(1999)]{1999LinkerJGR}Linker, J.A., Miki{\'c}, Z., Biesecker, D.A., Forsyth, R.J., Gibson, S.E., Lazarus, A.J., and, ...: 1999, {\it Journal of Geophysical Research} {\bf 104}, 9809. doi:10.1029/1998JA900159.
   
   
\bibitem[Lionello et al.(2013)]{2013LionelloApJ} Lionello, R., Downs, C., Linker, J.~A., et al.\ 2013, \apj, 777, 76. doi:10.1088/0004-637X/777/1/76


\bibitem[\protect\citeauthoryear{Lugaz et al.}{2010}]{2010LugazApJ} Lugaz N., Hernandez-Charpak J.~N., Roussev I.~I., Davis C.~J., Vourlidas A., Davies J.~A., 2010, ApJ, 715, 493. doi:10.1088/0004-637X/715/1/493


\bibitem[Lugaz \& Roussev(2011)]{2011LugazJASTP} Lugaz, N. \& Roussev, I.~I.\ 2011, Journal of Atmospheric and Solar-Terrestrial Physics, 73, 1187. doi:10.1016/j.jastp.2010.08.016

\bibitem[\protect\citeauthoryear{Lugaz et al.}{2020}]{2020LugazApJ} Lugaz N., Salman T.~M., Winslow R.~M., Al-Haddad N., Farrugia C.~J., Zhuang B., Galvin A.~B., 2020, ApJ, 899, 119. doi:10.3847/1538-4357/aba26b


\bibitem[\protect\citeauthoryear{Lynnyk et al.}{2011}]{2011Lynnyk} Lynnyk A., {\v{S}}afr{\'a}nkov{\'a} J., N{\v{e}}me{\v{c}}ek Z., Richardson J.~D., 2011, P\&SS, 59, 840. doi:10.1016/j.pss.2011.03.016

\bibitem[Maloney and Gallagher(2010)]{2010MaloneyApJ}Maloney, S.A. and Gallagher, P.T.: 2010, {\it The Astrophysical Journal} {\bf 724}, L127. doi:10.1088/2041-8205/724/2/L127.


\bibitem[\protect\citeauthoryear{Manchester et al.}{2004}]{2004Manchester} Manchester W.~B., Gombosi T.~I., Roussev I., de Zeeuw D.~L., Sokolov I.~V., Powell K.~G., T{\'o}th G., et al., 2004, JGRA, 109, A01102. doi:10.1029/2002JA009672

\bibitem[\protect\citeauthoryear{Marsch \& Tu}{1990}]{1990Marsch} Marsch E., Tu C.-Y., 1990, JGR, 95, 11945. doi:10.1029/JA095iA08p11945

\bibitem[Mishra and Wang(2018)]{2018MishraApJ}Mishra, W. and Wang, Y.: 2018, {\it The Astrophysical Journal} {\bf 865}, 50. doi:10.3847/1538-4357/aadb9b.


\bibitem[Mishra, Doshi, and Srivastava(2021)]{2021Mishra}Mishra, W., Doshi, U., and Srivastava, N.: 2021, {\it Frontiers in Astronomy and Space Sciences} {\bf 8}, 142. doi:10.3389/fspas.2021.713999.
         
      \bibitem[Moldwin \emph{et al.}(2000)]{2000MoldwinGeoRL}Moldwin, M.B., Ford, S., Lepping, R., Slavin, J., and Szabo, A.: 2000, {\it Geophys. Res. Lett.} {\bf 27}, 57. doi:10.1029/1999GL010724.

\bibitem[Moore(1959)]{1959MooreJFM}Moore, D.W.: 1959, {\it Journal of Fluid Mechanics} {\bf 6}, 113. doi:10.1017/S0022112059000520.


\bibitem[Moore(1963)]{1963MooreJFM}Moore, D.W.: 1963, {\it Journal of Fluid Mechanics} {\bf 16}, 161. doi:10.1017/S0022112063000665.


\bibitem[Nicolaou \emph{et al.}(2020)]{2020NicolaouApJ}Nicolaou, G., Livadiotis, G., Wicks, R.T., Verscharen, D., and Maruca, B.A.: 2020, {\it The Astrophysical Journal} {\bf 901}, 26. doi:10.3847/1538-4357/abaaae.

      
\bibitem[\protect\citeauthoryear{Nieves-Chinchilla et al.}{2012}]{2012NC} Nieves-Chinchilla T., Colaninno R., Vourlidas A., Szabo A., Lepping R.~P., Boardsen S.~A., Anderson B.~J., et al., 2012, JGRA, 117, A06106. doi:10.1029/2011JA017243

      
  \bibitem[Nieves-Chinchilla et al.(2016)]{2016NCApJ} Nieves-Chinchilla T., Linton M.~G., Hidalgo M.~A., Vourlidas A., Savani N.~P., Szabo A., Farrugia C., et al., 2016, ApJ, 823, 27. doi:10.3847/0004-637X/823/1/27


\bibitem[Nieves-Chinchilla et al.(2018)]{18NCSo} Nieves-Chinchilla, T., Vourlidas, A., Raymond, J.~C., et al.\ 2018, \solphys, 293, 25 
    
    \bibitem[Nieves-Chinchilla \emph{et al.}(2019)]{2019NCSoPh}Nieves-Chinchilla, T., Jian, L.K., Balmaceda, L., Vourlidas, A., dos Santos, L.F.G., and Szabo, A.: 2019, {\it Solar Physics} {\bf 294}, 89. doi:10.1007/s11207-019-1477-8.

\bibitem[Nieves-Chinchilla \emph{et al.}(2022)]{2022NCApJ}Nieves-Chinchilla, T., Alzate, N., Cremades, H., Rodr{\'\i}guez-Garc{\'\i}a, L., Dos Santos, L.F.G., Narock, A., and, ...: 2022, {\it The Astrophysical Journal} {\bf 930}, 88. doi:10.3847/1538-4357/ac590b.


\bibitem[\protect\citeauthoryear{Odstr{\v{c}}il \& Pizzo}{1999}]{1999OdstrcilJGR} Odstr{\v{c}}il D., Pizzo V.~J., 1999, JGR, 104, 493. doi:10.1029/1998JA900038
    

\bibitem[Odstrcil and Pizzo(2009)]{2009OdstrcilSoPh}Odstrcil, D. and Pizzo, V.J.: 2009, {\it Solar Physics} {\bf 259}, 297. doi:10.1007/s11207-009-9449-z.


 \bibitem[Parker(2009)]{09Pconf} Parker, E.~N.\ 2009, Climate and Weather of the Sun-earth System (CAWSES): Selected Papers from the 2007 Kyoto Symposium, 23
 
 \bibitem[Richardson and Cane(2010)]{2010RichCaneSoPh}Richardson, I.G. and Cane, H.V.: 2010, {\it Solar Physics} {\bf 264}, 189. doi:10.1007/s11207-010-9568-6.

 
 \bibitem[Rollett \emph{et al.}(2012)]{2012Rollett}Rollett, T., M{\"o}stl, C., Temmer, M., Veronig, A., and Farrugia, C.J.: 2012, {\it Solar Heliospheric and INterplanetary Environment (SHINE 2012)}.

      
   \bibitem[Russell et al.(2005)]{05RAdSpR} Russell, C.~T., Shinde, A.~A., \& Jian, L.\ 2005, Advances in Space Research, 35, 2178   


\bibitem[Sachdeva et al.(2015)]{15NApJ} Sachdeva, N., Subramanian, P., Colaninno, R., \& Vourlidas, A.\ 2015, \apj, 809, 158 

\bibitem[Sachdeva et al.(2017)]{17NSoPh} Sachdeva, N., Subramanian, P., Vourlidas, A., \& Bothmer, V.\ 2017, \solphys, 292, 118 

\bibitem[\protect\citeauthoryear{Savani et al.}{2010}]{2010SavaniApJL} Savani N.~P., Owens M.~J., Rouillard A.~P., Forsyth R.~J., Davies J.~A., 2010, ApJL, 714, L128. doi:10.1088/2041-8205/714/1/L128

      
\bibitem[Savani \emph{et al.}(2011)]{2011SavaniApJ}Savani, N.P., Owens, M.J., Rouillard, A.P., Forsyth, R.J., Kusano, K., Shiota, D., and, ...: 2011, {\it The Astrophysical Journal} {\bf 732}, 117. doi:10.1088/0004-637X/732/2/117.
  
\bibitem[Scolini \emph{et al.}(2019)]{2019ScoliniA&A}Scolini, C., Rodriguez, L., Mierla, M., Pomoell, J., and Poedts, S.: 2019, {\it Astronomy and Astrophysics} {\bf 626}, A122. doi:10.1051/0004-6361/201935053.
      
      
\bibitem[Spruit(2013)]{2013Spruit}Spruit, H.C.: 2013, {\it arXiv e-prints}, arXiv:1301.5572.
 
 \bibitem[\protect\citeauthoryear{St. Cyr et al.}{2000}]{2000StJGR} St. Cyr O.~C., Plunkett S.~P., Michels D.~J., Paswaters S.~E., Koomen M.~J., Simnett G.~M., Thompson B.~J., et al., 2000, JGR, 105, 18169. doi:10.1029/1999JA000381
 
   
\bibitem[T{\'o}th \emph{et al.}(2012)]{2012TothJCoPh}T{\'o}th, G., van der Holst, B., Sokolov, I.V., De Zeeuw, D.L., Gombosi, T.I., Fang, F., and, ...: 2012, {\it Journal of Computational Physics} {\bf 231}, 870. doi:10.1016/j.jcp.2011.02.006.
   
   \bibitem[Temmer(2021)]{2021TemmerLRSP}Temmer, M.: 2021, {\it Living Reviews in Solar Physics} {\bf 18}, 4. doi:10.1007/s41116-021-00030-3.

\bibitem[\protect\citeauthoryear{Verbeke et al.}{2022}]{2022Verbeke} Verbeke C., Schmieder B., D{\'e}moulin P., Dasso S., Grison B., Samara E., Scolini C., et al., 2022, arXiv, arXiv:2207.03168

   
\bibitem[von Steiger and Richardson(2006)]{2006vonSteigerSSRv}von Steiger, R. and Richardson, J.D.: 2006, {\it Space Science Reviews} {\bf 123}, 111. doi:10.1007/s11214-006-9015-z.
   
\bibitem[Vr{\v{s}}nak and Gopalswamy(2002)]{2002VrsnakJGRA}Vr{\v{s}}nak, B. and Gopalswamy, N.: 2002, {\it Journal of Geophysical Research (Space Physics)} {\bf 107}, 1019. doi:10.1029/2001JA000120.

   
\bibitem[Vr{\v{s}}nak et al.(2009)]{2009VrsIAUS} Vr{\v{s}}nak, B., Vrbanec, D., {\v{C}}alogovi{\'c}, J., et al.\ 2009, Universal Heliophysical Processes, 271   
   
   \bibitem[Vr{\v{s}}nak et al.(2013)]{2013VrsSoPh} Vr{\v{s}}nak, B., {\v{Z}}ic, T., Vrbanec, D., et al.\ 2013, \solphys, 285, 295   
   
   
   

\bibitem[\protect\citeauthoryear{Wang \& Richardson}{2004}]{2004WangJGRA} Wang C., Richardson J.~D., 2004, JGRA, 109, A06104. doi:10.1029/2004JA010379

\bibitem[\protect\citeauthoryear{Webb et al.}{2009}]{2009WebbSoPh} Webb D.~F., Howard T.~A., Fry C.~D., Kuchar T.~A., Odstrcil D., Jackson B.~V., Bisi M.~M., et al., 2009, SoPh, 256, 239. doi:10.1007/s11207-009-9351-8

\bibitem[Weber and Davis(1967)]{1967WeberApJ}Weber, E.J. and Davis, L.: 1967, {\it The Astrophysical Journal} {\bf 148}, 217. doi:10.1086/149138.


\bibitem[Weiss \emph{et al.}(2020)]{2020WeissEGUGA}Weiss, A., M{\"o}stl, C., Nieves-Chinchilla, T., Amerstorfer, T., Palmerio, E., Reiss, M., and, ...: 2020, {\it EGU General Assembly Conference Abstracts}. doi:10.5194/egusphere-egu2020-8398.

    
\bibitem[Xie, Ofman, and Lawrence(2004)]{2004XieJGRA}Xie, H., Ofman, L., and Lawrence, G.: 2004, {\it Journal of Geophysical Research (Space Physics)} {\bf 109}, A03109. doi:10.1029/2003JA010226.
    
      
\end{thebibliography}
%

\begin{appendix} 
	\section{Data Table}
	
	\begin{landscape}
		\begin{table}[h]

			\caption{
				The list of the 152 WIND ICME events we use in this study. The arrival date and time of the ICME at the position of WIND measurement and the arrival and departure dates \& times of the associated magnetic clouds (MCs) are taken from WIND ICME catalogue (\url{https://wind.nasa.gov/ICMEindex.php}). The 14 events marked with and asterisk (*) coincide with the near earth counterparts of 14 CMEs listed in \citet{17NSoPh}. }  
			\begin{center}
				
				\begin{tabular}{cccclccc}
					
					\hline
					\hline
					CME        & CME Arrival date & MC start  & MC end       & Flux rope \\
					event      & and time[UT]   &  date and   & date and     & type   \\
					number    &  (1AU)         &  time [UT]   &   time [UT]   &      \\
					\hline
					
					1    &  1995 03 04 , 00:36 & 1995 03 04 , 11:23 & 1995 03 05 , 03:06 & Fr \\
					2    &  1995 04 03 , 06:43 & 1995 04 03 , 12:45 & 1995 04 04 , 13:25 & F+ \\
					3    &  1995 06 30 , 09:21 & 1995 06 30 , 14:23 & 1995 07 02 , 16:47 & Fr \\
					4    &  1995 08 22 , 12:56 & 1995 08 22 , 22:19 & 1995 08 23 , 18:43 & Fr \\
					5    &  1995 09 26 , 15:57 & 1995 09 27 , 03:36 & 1995 09 27 , 21:21 & Fr \\
					6    &  1995 10 18 , 10:40 & 1995 10 18 , 19:11 & 1995 10 20 , 02:23 & Fr \\
					7    &  1996 02 15 , 15:07 & 1996 02 15 , 15:07 & 1996 02 16 , 08:59 & F+ \\
					8    &  1996 04 04 , 11:59 & 1996 04 04 , 11:59 & 1996 04 04 , 21:36 & Fr \\ 
					9    &  1996 05 16 , 22:47 & 1996 05 17 , 01:36 & 1996 05 17 , 11:58 & F+ \\
					10   &  1996 05 27 , 14:45 & 1996 05 27 , 14:45 & 1996 05 29 , 02:22 & Fr \\
					11   &  1996 07 01 , 13:05 & 1996 07 01 , 17:16 & 1996 07 02 , 10:17 & Fr \\
					12   &  1996 08 07 , 08:23 & 1996 08 07 , 11:59 & 1996 08 08 , 13:12 & Fr \\
					13   &  1996 12 24 , 01:26 & 1996 12 24 , 03:07 & 1996 12 25 , 11:44 & F+ \\ 
					14   &  1997 01 10 , 00:52 & 1997 01 10 , 04:47 & 1997 01 11 , 03:36 & F+ \\
					15   &  1997 04 10 , 17:02 & 1997 04 11 , 05:45 & 1997 04 11 , 19:10 & Fr \\
					16   &  1997 04 21 , 10:11 & 1997 04 21 , 11:59 & 1997 04 23 , 07:11 & F+ \\
					17   &  1997 05 15 , 01:15 & 1997 05 15 , 10:00 & 1997 05 16 , 02:37 & F+ \\
					18   &  1997 05 26 , 09:09 & 1997 05 26 , 15:35 & 1997 05 28 , 00:00 & Fr \\
					19   &  1997 06 08 , 15:43 & 1997 06 09 , 06:18 & 1997 06 09 , 23:01 & Fr \\
					20   &  1997 06 19 , 00:00 & 1997 06 19 , 05:31 & 1997 06 20 , 22:29 & Fr \\
					21   &  1997 07 15 , 03:10 & 1997 07 15 , 06:48 & 1997 07 16 , 11:16 & F+ \\
					22   &  1997 08 03 , 10:10 & 1997 08 03 , 13:55 & 1997 08 04 , 02:23 & Fr \\
					23   &  1997 08 17 , 01:56 & 1997 08 17 , 06:33 & 1997 08 17 , 20:09 & Fr \\
					24   &  1997 09 02 , 22:40 & 1997 09 03 , 08:38 & 1997 09 03 , 20:59 & Fr \\
					25   &  1997 09 18 , 00:30 & 1997 09 18 , 04:07 & 1997 09 19 , 23:59 & F+ \\
					26   &  1997 10 01 , 11:45 & 1997 10 01 , 17:08 & 1997 10 02 , 23:15 & Fr \\
					27   &  1997 10 10 , 03:08 & 1997 10 10 , 15:33 & 1997 10 11 , 22:00 & F+ \\
					28   &  1997 11 06 , 22:25 & 1997 11 07 , 06:00 & 1997 11 08 , 22:46 & F+ \\
					29   &  1997 11 22 , 09:12 & 1997 11 22 , 17:31 & 1997 11 23 , 18:43 & F+ \\
					30   &  1997 12 30 , 01:13 & 1997 12 30 , 09:35 & 1997 12 31 , 08:51 & Fr \\
					31   &  1998 01 06 , 13:29 & 1998 01 07 , 02:23 & 1998 01 08 , 07:54 & F+ \\
					32   &  1998 01 28 , 16:04 & 1998 01 29 , 13:12 & 1998 01 31 , 00:00 & F+ \\
					33   &  1998 03 25 , 10:48 & 1998 03 25 , 14:23 & 1998 03 26 , 08:57 & Fr \\
					34   &  1998 03 31 , 07:11 & 1998 03 31 , 11:59 & 1998 04 01 , 16:18 & Fr \\
					35   &  1998 05 01 , 21:21 & 1998 05 02 , 11:31 & 1998 05 03 , 16:47 & Fr \\
					36   &  1998 06 02 , 10:28 & 1998 06 02 , 10:28 & 1998 06 02 , 09:16 & Fr \\
					37   &  1998 06 24 , 10:47 & 1998 06 24 , 13:26 & 1998 06 25 , 22:33 & F+ \\
					38   &  1998 07 10 , 22:36 & 1998 07 10 , 22:36 & 1998 07 12 , 21:34 & F+ \\
					39   &  1998 08 19 , 18:40 & 1998 08 20 , 08:38 & 1998 08 21 , 20:09 & F+ \\
					40   &  1998 10 18 , 19:30 & 1998 10 19 , 04:19 & 1998 10 20 , 07:11 & F+ \\
					
					\hline
					
				\end{tabular}
			\end{center}
		\end{table}
		
	\end{landscape}

	\begin{landscape} 
		\begin{table}
			\caption{continued}
			\begin{center}
				\begin{tabular}{cccclccc}
					\hline
					\hline
					CME        & CME Arrival date & MC start  & MC end       & Flux rope \\
					event      & and time[UT]   &  date and   & date and     & type   \\
					number    &  (1AU)         &  time [UT]   &   time [UT]   &      \\
					\hline
					
					41   &  1999 02 11 , 17:41 & 1999 02 11 , 17:41 & 1999 02 12 , 03:35 & Fr \\
					42   &  1999 07 02 , 00:27 & 1999 07 03 , 08:09 & 1999 07 05 , 13:13 & Fr \\
					43   &  1999 09 21 , 18:57 & 1999 09 21 , 18:57 & 1999 09 22 , 11:31 & Fr \\
					44   &  2000 02 11 , 23:34 & 2000 02 12 , 12:20 & 2000 02 13 , 00:35 & Fr \\
					45   &  2000 02 20 , 21:03 & 2000 02 21 , 14:24 & 2000 02 22 , 13:16 & Fr \\

					46   &  2000 03 01 , 01:58 & 2000 03 01 , 03:21 & 2000 03 02 , 03:07 & Fr \\
					47   &  2000 07 01 , 07:12 & 2000 07 01 , 07:12 & 2000 07 02 , 03:34 & Fr \\
					48   &  2000 07 11 , 22:35 & 2000 07 11 , 22:35 & 2000 07 13 , 04:33 & Fr \\
					49   &  2000 07 28 , 06:38 & 2000 07 28 , 14:24 & 2000 07 29 , 10:06 & F+ \\
					50   &  2000 09 02 , 23:16 & 2000 09 02 , 23:16 & 2000 09 03 , 22:32 & Fr \\
					51   &  2000 10 03 , 01:02 & 2000 10 03 , 09:36 & 2000 10 05 , 03:34 & F+ \\
					52   &  2000 10 12 , 22:33 & 2000 10 13 , 18:24 & 2000 10 14 , 19:12 & Fr \\
					53   &  2000 11 06 , 09:30 & 2000 11 06 , 23:05 & 2000 11 07 , 18:05 & Fr \\
					54   &  2000 11 26 , 11:43 & 2000 11 27 , 09:30 & 2000 11 28 , 09:36 & Fr \\
					55   &  2001 04 21 , 15:29 & 2001 04 22 , 00:28 & 2001 04 23 , 01:11 & Fr \\
					56   &  2001 10 21 , 16:39 & 2001 10 22 , 01:17 & 2001 10 23 , 00:47 & Fr \\
					57   &  2001 11 24 , 05:51 & 2001 11 24 , 15:47 & 2001 11 25 , 13:17 & Fr \\
					58   &  2001 12 29 , 05:16 & 2001 12 30 , 03:24 & 2001 12 30 , 19:10 & Fr \\
					59   &  2002 02 28 , 05:06 & 2002 02 28 , 19:11 & 2002 03 01 , 23:15 & Fr \\
					60   &  2002 03 18 , 13:14 & 2002 03 19 , 06:14 & 2002 03 20 , 15:36 & Fr \\
					61   &  2002 03 23 , 11:24 & 2002 03 24 , 13:11 & 2002 03 25 , 21:36 & Fr \\
					62   &  2002 04 17 , 11:01 & 2002 04 17 , 21:36 & 2002 04 19 , 08:22 & F+ \\
					63   &  2002 07 17 , 15:56 & 2002 07 18 , 13:26 & 2002 07 19 , 09:35 & Fr \\
					64   &  2002 08 18 , 18:40 & 2002 08 19 , 19:12 & 2002 08 21 , 13:25 & Fr \\
					65   &  2002 08 26 , 11:16 & 2002 08 26 , 14:23 & 2002 08 27 , 10:47 & Fr \\
					66   &  2002 09 30 , 07:54 & 2002 09 30 , 22:04 & 2002 10 01 , 20:08 & F+ \\
					67   &  2002 12 21 , 03:21 & 2002 12 21 , 10:20 & 2002 12 22 , 15:36 & Fr \\
					68   &  2003 01 26 , 21:43 & 2003 01 27 , 01:40 & 2003 01 27 , 16:04 & Fr \\
					69   &  2003 02 01 , 13:06 & 2003 02 02 , 19:11 & 2003 02 03 , 09:35 & Fr \\
					70   &  2003 03 20 , 04:30 & 2003 03 20 , 11:54 & 2003 03 20 , 22:22 & Fr \\
					71   &  2003 06 16 , 22:33 & 2003 06 16 , 17:48 & 2003 06 18 , 08:18 & Fr \\
					72   &  2003 08 04 , 20:23 & 2003 08 05 , 01:10 & 2003 08 06 , 02:23 & Fr \\
					73   &  2003 11 20 , 08:35 & 2003 11 20 , 11:31 & 2003 11 21 , 01:40 & Fr \\
					74   &  2004 04 03 , 09:55 & 2004 04 04 , 01:11 & 2004 04 05 , 19:11 & F+ \\
					75   &  2004 09 17 , 20:52 & 2004 09 18 , 12:28 & 2004 09 19 , 16:58 & Fr \\
					
					76   &  2005 05 15 , 02:10 & 2005 05 15 , 05:31 & 2005 05 16 , 22:47 & F+ \\
					77   &  2005 05 20 , 04:47 & 2005 05 20 , 09:35 & 2005 05 22 , 02:23 & F+ \\
					78   &  2005 07 17 , 14:52 & 2005 07 17 , 14:52 & 2005 07 18 , 05:59 & Fr \\
					79   &  2005 10 31 , 02:23 & 2005 10 31 , 02:23 & 2005 10 31 , 18:42 & Fr \\
					80   &  2006 02 05 , 18:14 & 2006 02 05 , 20:23 & 2006 02 06 , 11:59 & F+ \\
					81   &  2006 09 30 , 02:52 & 2006 09 30 , 08:23 & 2006 09 30 , 22:03 & F+ \\
					82   &  2006 11 18 , 07:11 & 2006 11 18 , 07:11 & 2006 11 20 , 04:47 & Fr \\
					83   &  2007 05 21 , 22:40 & 2007 05 21 , 22:45 & 2007 05 22 , 13:25 & Fr \\
					84   &  2007 06 08 , 05:45 & 2007 06 08 , 05:45 & 2007 06 09 , 05:15 & Fr \\
					85   &  2007 11 19 , 17:22 & 2007 11 20 , 00:33 & 2007 11 20 , 11:31 & Fr \\
					
					\hline
					
				\end{tabular}
			\end{center}
		\end{table}
	\end{landscape}

	\begin{landscape} 
		
		\begin{table}
			\caption{continued}
			\begin{center}
				\begin{tabular}{cccclccc}
					\hline
					\hline
					CME        & CME Arrival date & MC start  & MC end       & Flux rope \\
					event      & and time[UT]   &  date and   & date and     & type   \\
					number    &  (1AU)         &  time [UT]   &   time [UT]   &      \\
					\hline
					
					86   &  2008 05 23 , 01:12 & 2008 05 23 , 01:12 & 2008 05 23 , 10:46 & F+ \\
					87   &  2008 09 03 , 16:33 & 2008 09 03 , 16:33 & 2008 09 04 , 03:49 & F+ \\
					88   &  2008 09 17 , 00:43 & 2008 09 17 , 03:57 & 2008 09 18 , 08:09 & Fr \\
					89   &  2008 12 04 , 11:59 & 2008 12 04 , 16:47 & 2008 12 05 , 10:47 & Fr \\
					90   &  2008 12 17 , 03:35 & 2008 12 17 , 03:35 & 2008 12 17 , 15:35 & Fr \\
					91   &  2009 02 03 , 19:21 & 2009 02 03 , 01:12 & 2009 02 04 , 19:40 & F+ \\
					92   &  2009 03 11 , 22:04 & 2009 03 12 , 01:12 & 2009 03 13 , 01:40 & F+ \\
					93   &  2009 04 22 , 11:16 & 2009 04 22 , 14:09 & 2009 04 22 , 20:37 & Fr \\
					94   &  2009 06 03 , 13:40 & 2009 06 03 , 20:52 & 2009 06 05 , 05:31 & Fr \\
					95   &  2009 06 27 , 11:02 & 2009 06 27 , 17:59 & 2009 06 28 , 20:24 & F+ \\

					96   &  2009 07 21 , 02:53 & 2009 07 21 , 04:48 & 2009 07 22 , 03:36 & Fr \\
					97   &  2009 09 10 , 10:19 & 2009 09 10 , 10:19 & 2009 09 10 , 19:26 & Fr \\
					
					98   &  2009 09 30 , 00:44 & 2009 09 30 , 06:59 & 2009 09 30 , 19:11 & Fr \\
					99   &  2009 10 29 , 01:26 & 2009 10 29 , 01:26 & 2009 10 29 , 23:45 & F+ \\
					100   &  2009 11 14 , 10:47 & 2009 11 14 , 10:47 & 2009 11 15 , 11:45 & Fr \\
					101   &  2009 12 12 , 04:47 & 2009 12 12 , 19:26 & 2009 12 14 , 04:47 & Fr \\
					102   &  2010 01 01 , 22:04 & 2010 01 02 , 00:14 & 2010 01 03 , 09:06 & Fr \\
					103   &  2010 02 07 , 18:04 & 2010 02 07 , 19:11 & 2010 02 09 , 05:42 & Fr \\
					104*   &  2010 03 23 , 22:29 & 2010 03 23 , 22:23 & 2010 03 24 , 15:36 & Fr \\
					105*   &  2010 04 05 , 07:55 & 2010 04 05 , 11:59 & 2010 04 06 , 16:48 & Fr \\
					106*   &  2010 04 11 , 12:20 & 2010 04 11 , 21:36 & 2010 04 12 , 14:12 & Fr \\
					107   &  2010 05 28 , 01:55 & 2010 05 29 , 19:12 & 2010 05 29 , 17:58 & Fr \\
					108*   &  2010 06 21 , 03:35 & 2010 06 21 , 06:28 & 2010 06 22 , 12:43 & Fr \\
					109*   &  2010 09 15 , 02:24 & 2010 09 15 , 02:24 & 2010 09 16 , 11:58 & Fr \\
					110*   &  2010 10 31 , 02:09 & 2010 10 30 , 05:16 & 2010 11 01 , 20:38 & Fr \\
					111   &  2010 12 19 , 00:35 & 2010 12 19 , 22:33 & 2010 12 20 , 22:14 & F+ \\
					112   &  2011 01 24 , 06:43 & 2011 01 24 , 10:33 & 2011 01 25 , 22:04 & F+ \\
					113*   &  2011 03 29 , 15:12 & 2011 03 29 , 23:59 & 2011 04 01 , 14:52 & Fr \\
					114   &  2011 05 28 , 00:14 & 2011 05 28 , 05:31 & 2011 05 28 , 22:47 & F+ \\
					115   &  2011 06 04 , 20:06 & 2011 06 05 , 01:12 & 2011 06 05 , 18:13 & Fr \\
					116   &  2011 07 03 , 19:12 & 2011 07 03 , 19:12 & 2011 07 04 , 19:12 & Fr \\
					117*   &  2011 09 17 , 02:57 & 2011 09 17 , 15:35 & 2011 09 18 , 21:07 & Fr \\
					118   &  2012 02 14 , 07:11 & 2012 02 14 , 20:52 & 2012 02 16 , 04:47 & Fr \\
					119   &  2012 04 05 , 14:23 & 2012 04 05 , 19:41 & 2012 04 06 , 21:36 & Fr \\
					120   &  2012 05 03 , 00:59 & 2012 05 04 , 03:36 & 2012 05 05 , 11:22 & Fr \\
					121   &  2012 05 16 , 12:28 & 2012 05 16 , 16:04 & 2012 05 18 , 02:11 & Fr \\
					122   &  2012 06 11 , 02:52 & 2012 06 11 , 11:31 & 2012 06 12 , 05:16 & Fr \\
					123*   &  2012 06 16 , 09:03 & 2012 06 16 , 22:01 & 2012 06 17 , 11:23 & F+ \\
					124*   &  2012 07 14 , 17:39 & 2012 07 15 , 06:14 & 2012 07 17 , 03:22 & Fr \\
					125   &  2012 08 12 , 12:37 & 2012 08 12 , 19:12 & 2012 08 13 , 05:01 & Fr \\
					126   &  2012 08 18 , 03:25 & 2012 08 18 , 19:12 & 2012 08 19 , 08:22 & Fr \\
					127*   &  2012 10 08 , 04:12 & 2012 10 08 , 15:50 & 2012 10 09 , 17:17 & Fr \\
					128   &  2012 10 12 , 08:09 & 2012 10 12 , 18:09 & 2012 10 13 , 09:14 & Fr \\
					129*   &  2012 10 31 , 14:28 & 2012 10 31 , 23:35 & 2012 11 02 , 05:21 & F+ \\
					130*   &  2013 03 17 , 05:21 & 2013 03 17 , 14:09 & 2013 03 19 , 16:04 & Fr \\

					\hline
					
				\end{tabular}
			\end{center}
		\end{table}
		
	\end{landscape}
	
	\begin{landscape} 
		\begin{table}
			\caption{continued}
			\begin{center}
				\begin{tabular}{cccclccc}
					\hline
					\hline
					CME        & CME Arrival date & MC start  & MC end       & Flux rope \\
					event      & and time[UT]   &  date and   & date and     & type   \\
					number    &  (1AU)         &  time [UT]   &   time [UT]   &      \\
					\hline

					
					131*   &  2013 04 13 , 22:13 & 2013 04 14 , 17:02 & 2013 04 17 , 05:30 & F+ \\
					132   &  2013 04 30 , 08:52 & 2013 04 30 , 12:00 & 2013 05 01 , 07:12 & Fr \\
					133   &  2013 05 14 , 02:23 & 2013 05 14 , 06:00 & 2013 05 15 , 06:28 & Fr \\
					134   &  2013 06 06 , 02:09 & 2013 06 06 , 14:23 & 2013 06 08 , 00:00 & F+ \\
					135   &  2013 06 27 , 13:51 & 2013 06 28 , 02:23 & 2013 06 29 , 11:59 & Fr \\
					136   &  2013 09 01 , 06:14 & 2013 09 01 , 13:55 & 2013 09 02 , 01:56 & Fr \\
					137   &  2013 10 30 , 18:14 & 2013 10 30 , 18:14 & 2013 10 31 , 05:30 & Fr \\
					138   &  2013 11 08 , 21:07 & 2013 11 08 , 23:59 & 2013 11 09 , 06:14 & Fr \\
					139   &  2013 11 23 , 00:14 & 2013 11 23 , 04:47 & 2013 11 23 , 15:35 & Fr \\
					140   &  2013 12 14 , 16:47 & 2013 12 15 , 16:47 & 2013 12 16 , 05:30 & Fr \\

					141   &  2013 12 24 , 20:36 & 2013 12 25 , 04:47 & 2013 12 25 , 17:59 & F+ \\
					142   &  2014 04 05 , 09:58 & 2014 04 05 , 22:18 & 2014 04 07 , 14:24 & Fr \\
					143   &  2014 04 11 , 06:57 & 2014 04 11 , 06:57 & 2014 04 12 , 20:52 & F+ \\
					144   &  2014 04 20 , 10:20 & 2014 04 21 , 07:41 & 2014 04 22 , 06:12 & Fr \\
					145   &  2014 04 29 , 19:11 & 2014 04 29 , 19:11 & 2014 04 30 , 16:33 & Fr \\
					146   &  2014 06 29 , 04:47 & 2014 06 29 , 20:53 & 2014 06 30 , 11:15 & Fr \\
					147   &  2014 08 19 , 05:49 & 2014 08 19 , 17:59 & 2014 08 21 , 19:09 & F+ \\
					148   &  2014 08 26 , 02:40 & 2014 08 27 , 03:07 & 2014 08 27 , 21:49 & Fr \\
					149   &  2015 01 07 , 05:38 & 2015 01 07 , 06:28 & 2015 01 07 , 21:07 & F+ \\
					150   &  2015 09 07 , 13:05 & 2015 09 07 , 23:31 & 2015 09 09 , 14:52 & F+ \\
					151   &  2015 10 06 , 21:35 & 2015 10 06 , 21:35 & 2015 10 07 , 10:03 & Fr \\
					152   &  2015 12 19 , 15:35 & 2015 12 20 , 13:40 & 2015 12 21 , 23:02 & Fr \\
					\hline
					
				\end{tabular}
			\end{center}
		\end{table}      
	\end{landscape}
	\label{S - Table A}

	\newpage
	
	\section{The List of $C_{\rm x}$, $C_{\rm p}$ and $C_{\rm r}$ for All The Events in Our Dataset}
	\newpage
	\begin{landscape}
		\begin{table}[h!]

			\caption{
				$C_{\rm x}$ (Equation~\ref{eq: Cx}), $C_{\rm p}$ (Equation~\ref{eq: Cp}) and $C_{\rm r}$ (Equation~\ref{eq: Cr}) for all the events in the dataset. The 14 events marked with and asterisk (*) coincide with the near earth counterparts of 14 CMEs listed in \citet{17NSoPh}. }  
			\begin{center}
				
				\begin{tabular}{|c|ccc|ccc|c|}
					
					\hline
					\hline
					CME        & \multicolumn{3}{c|}{Using BG1} &  \multicolumn{3}{c|}{Using BG2} \\ 
					event      &  $C_{\rm x}$  & $C_{\rm p}$    &  $C_{\rm r}$ &  $C_{\rm x}$  & $C_{\rm p}$    &  $C_{\rm r}$    \\
					number    &    &    &    &   &   &    \\
					\hline
					
					1 &	2.11 &	6.85 &	1.78 &	28.08 &	6.89 &	3.34	\\ 
					2 &	24.93 &	3.23 &	0.52 &	5.71 &	0.91 &	2.95 \\ 
					3 &	6.35 &	1.82 &	3.47 &	7.62 &	1.42 &	4.16 \\ 
					4 &	9.36 &	2.68 &	2.29 &	19.19 &	6.56 &	8.23 \\ 
					5 &	11.35 &	4.14 &	7.55 &	7.09 &	4.42 &	2.52 \\ 
					6 &	69.85 &	16.07 &	15.37 &	42.55 &	11.20 &	7.23 \\ 
					7 &	0.49 &	1.65 &	-0.20 &	5.68 &	1.80 &	0.26 \\ 
					8 &	4.79 &	1.74 &	3.22 &	4.94 &	1.82 &	10.74 \\ 
					9 &	1.16 &	1.62 &	0.29 &	13.06 &	2.55 &	18.30 \\ 
					10 & 7.35 &	2.63 &	2.17 &	13.03 &	4.78 &	4.13 \\ 
					11 & 5.65 &	2.90 &	5.80 &	10.37 &	7.06 &	4.82 \\ 
					12 & 6.53 &	1.70 &	2.36 &	5.08 &	1.72 &	4.80 \\ 
					13 & 8.77 &	1.96 &	1.70 &	7.80 &	2.34 &	2.38 \\ 
					14 & 18.79 & 4.08 &	18.01 &	40.73 &	8.69 &	8.38 \\ 
					15 & 5.62 &	7.52 &	7.08 &	13.68 &	5.35 &	1.06 \\ 
					16 & 3.94 &	12.08 &	3.48 &	15.62 &	7.96 &	9.40 \\ 
					17 & 83.94 & 24.15 & 1.88 &	32.90 &	7.72 &	2.01 \\ 
					18 & 4.69 &	2.14 &	24.73 &	4.06 &	1.74 &	4.50 \\ 
					19 & 37.80 & 4.98 &	11.83 &	22.26 &	4.01 &	30.68 \\ 
					20 & 11.57 & 3.84 &	15.12 &	11.57 &	3.84 &	15.12 \\ 
					21 & 13.06 & 3.61 &	9.08 &	12.89 &	5.15 &	4.86 \\ 
					22 & 7.01 &	6.34 &	10.06 &	34.91 &	5.75 &	12.53 \\ 
					23 & 7.07 &	1.35 &	4.12 &	7.63 &	1.41 &	4.87 \\ 
					24 & 7.87 &	10.35 &	9.93 &	18.73 &	10.51 &	2.35 \\ 
					25 & 4.23 &	7.82 &	15.08 &	7.60 &	4.65 &	13.96 \\ 
					26 & 20.23 & 2.32 &	10.74 &	2.24 &	0.15 &	0.30 \\ 
					27 & 10.62 & 2.78 &	35.34 &	11.12 &	2.89 &	30.14 \\ 
					28 & 21.81 & 8.47 &	3.23 &	15.95 &	3.71 &	4.32 \\ 
					29 & 46.27 & 8.62 &	1.95 &	21.75 &	5.84 &	1.66 \\ 
					30 & 30.85 & 14.60 & 4.16 &	21.60 &	5.63 &	7.22 \\ 
					31 & 34.43 & 8.88 &	21.85 &	17.12 &	6.18 &	4.05 \\ 
					32 & 10.44 & 1.60 &	1.57 &	15.60 &	2.73 &	3.71 \\ 
					33 & 17.38 & 4.91 &	1.03 &	5.52 &	2.27 &	4.65 \\ 
					34 & 1.06 &	1.06 &	0.01 &	0.96 &	1.10 &	-0.03 \\ 
					35 & 8.15 &	4.37 &	0.16 &	3.41 &	1.18 &	0.31 \\ 
					
					36 & 2.88 &	3.05 &	2.15 &	13.91 &	3.07 &	4.65 \\ 
					37 & 1.97 &	4.01 &	1.72 &	5.76 &	1.66 &	1.07 \\ 
					38 & 40.72 & 6.69 &	50.49 &	16.17 &  2.24 &	6.40 \\ 
					39 & 11.18 & 3.60 &	3.26 &	20.53 &	6.48 &	9.92 \\ 
					40 & 17.51 & 6.37 &	6.66 &	42.52 &	5.20 &	14.20 \\ 

					\hline
					
				\end{tabular}
			\end{center}
		\end{table}
		
	\end{landscape}

	\begin{landscape} 
		\begin{table}
			\caption{continued}
			\begin{center}
				\begin{tabular}{|c|ccc|ccc|c|}
					
					\hline
					\hline
					CME        & \multicolumn{3}{c|}{Using BG1} &  \multicolumn{3}{c|}{Using BG2} \\ 
					event      &  $C_{\rm x}$  & $C_{\rm p}$    &  $C_{\rm r}$ &  $C_{\rm x}$  & $C_{\rm p}$    &  $C_{\rm r}$    \\
					number    &    &    &    &   &   &    \\
					\hline
					
					41 & 10.79 & 8.44 &	7.76 &	4.74 &	5.71 &	1.43 \\ 
					42 & 29.70 & 0.82 &	1.15 &	17.42 &	0.67 &	1.63 \\ 
					43 & 1.56 &	3.52 &	3.96 &	4.12 &	2.98 &	9.42 \\ 
					44 & 1.94 &	5.38 &	2.80 &	9.30 &	1.81 &	1.30 \\ 
					45 & 38.53 & 4.29 &	19.14 &	23.95 &	3.40 &	12.96 \\ 
					46 & 19.02 & 2.69 &	1.07 & 18.73 &	2.68 &	1.20 \\ 
					47 & 1.18 &	2.83 &	0.69 &	3.79 &	1.74 &	1.85 \\ 
					48 & 5.04 &	2.35 &	1.77 & 	3.55 &	0.23 &	1.52 \\ 
					49 & 10.12 & 4.39 &	1.01 &	9.48 &	5.30 &	1.04 \\ 
					50 & 10.11 & 1.36 &	0.32 &	4.08 &	0.63 &	0.69 \\ 
					51 & 17.47 & 6.91 &	6.55 &	10.04 &	4.95 &  9.27 \\ 
					52 & 5.09 & 2.81 &	2.33 &	17.54 &	3.37 &	4.06 \\ 
					53 & 31.83 & 18.57 & 3.61 &	87.69 &	5.66 &	14.80 \\ 
					54 & 8.37 &	3.36 &	0.16 &	7.67 &	1.71 &	0.33 \\ 
					55 & 18.96 & 6.35 &	9.46 &	28.15 &	8.26 &	24.59 \\ 
					56 & 1.77 &	3.58 &	0.04 &	1.68 &	3.50 &	0.04 \\ 
					57 & 189.13 & 7.05 & 1.40 &	177.90 & 4.06 &	2.15 \\ 
					58 & 42.62 & 5.95 &	60.55 &	17.51 &	4.67 &	32.86 \\ 
					59 & 4.05 &	2.04 &	1.65 &	4.75 &	2.29 &	1.35 \\ 
					60 & 133.84 & 5.51 & 22.01 & 93.76 & 6.20 &	19.50 \\ 
					61 & 41.65 & 3.89 &	47.33 &	84.49 &	5.60 &	34.64 \\ 
					62 & 44.67 & 2.48 &	3.51 &	38.32 &	2.01 &	3.27 \\ 
					63 & 116.44 & 3.04 & 13.85 & 49.05 & 1.29 &	68.57 \\ 
					64 & 18.86 & 0.81 &	1.25 &	21.38 &	1.94 &	3.73 \\ 
					65 & 3.16 &	4.67 &	1.05 &	1.44 &	3.45 &	0.17 \\ 
					66 & 17.37 & 10.03 & 11.50 & 5.07 &	3.36 &	3.74 \\ 
					67 & 5.90 &	1.86 &	1.95 &	4.66 &	1.58 &	0.57 \\ 
					68 & 48.21 & 1.64 &	0.86 &	34.78 &	1.71 &	1.43 \\ 
					69 & 6.77 &	0.85 &	1.05 &	9.15 &	1.76 &	0.24 \\ 
					70 & 41.44 & 1.56 &	13.90 &	39.08 &	1.57 &	8.93 \\ 
					71 & 55.44 & 7.18 &	9.34 &	7.72 &	1.48 &	5.16 \\ 
					72 & 7.63 &	2.71 &	0.04 &	4.49 &	3.29 &	0.32 \\ 
					73 & 299.67 & 35.64 & 12.01 & 211.51 & 38.28 &  8.27 \\ 
					74 & 42.12 & 8.24 &	27.90 &	27.12 &	6.92 &	9.22 \\ 
					75 & 2.31 &	0.88 &	0.06 &	1.46 &	0.65 &	0.06 \\ 
					76 & 228.01 & 19.54 & 2.54 & 252.44 & 21.85 & 2.32 \\ 
					77 & 10.97 & 1.75 &	1.43 &	19.42 &	3.18 &	5.24 \\ 
					78 & 23.28 & 9.95 &	4.86 &	11.68 &	5.32 &	2.52 \\ 
					79 & 14.42 & 6.83 &	14.05 &	6.63 &	3.75 &	3.17 \\ 
					80 & 7.29 &	2.46 &	5.57 &	5.97 &	2.35 &	8.63 \\ 
					81 & 44.21 & 18.40 & 2.87 &	15.35 &	13.98 &	1.49 \\ 
					82 & 7.80 &	1.88 &	13.10 &	58.77 &	2.01 &	21.42 \\ 
					83 & 42.13 & 6.85 &	0.85 &	23.76 &	6.67 &	3.49 \\ 
					84 & 23.96 & 6.14 &	12.48 &	7.88 &	2.81 &	7.85 \\ 
					85 & 78.21 & 18.09 & 3.95 &	58.24 &	12.72 &	3.60 \\ 

					\hline
					
				\end{tabular}
			\end{center}
		\end{table}
	\end{landscape}

	\begin{landscape} 
		
		\begin{table}
			\caption{continued}
			\begin{center}
				\begin{tabular}{|c|ccc|ccc|c|}
					
					\hline
					\hline
					CME        & \multicolumn{3}{c|}{Using BG1} &  \multicolumn{3}{c|}{Using BG2} \\ 
					event      &  $C_{\rm x}$  & $C_{\rm p}$    &  $C_{\rm r}$ &  $C_{\rm x}$  & $C_{\rm p}$    &  $C_{\rm r}$    \\
					number    &    &    &    &   &   &    \\
					\hline
					
					86 & 1.28 &	1.33 &	0.36 &	18.99 &	1.31 &	1.76 \\ 
					87 & 20.65 & 7.36 &	0.59 &	6.09 &	1.44 &	0.86 \\ 
					88 & 21.97 & 2.14 &	0.54 &	22.01 &	2.17 &	0.55 \\ 
					89 & 21.43 & 5.62 &	1.03 &	7.17 &	0.76 &	1.26 \\ 
					90 & 57.20 & 4.35 &	61.71 &	22.71 &	1.63 &	34.01 \\ 
					91 & 29.19 & 8.97 &	7.29 &	15.22 &	12.15 &	6.30 \\ 
					92 & 57.55 & 11.76 & 15.98 & 19.02 & 9.73 &	8.38 \\ 
					93 & 0.97 &	0.78 &	-0.03 &	14.13 &	1.14 &	6.47 \\ 
					94 & 6.72 &	1.89 &	10.43 &	6.07 &	1.98 &	7.64 \\ 
					95 & 33.10 & 5.15 &	4.08 &	25.83 &	5.18 &	16.91 \\ 
					96 & 6.92 &	10.10 &	15.12 &	4.12 &	1.47 &	6.78 \\ 
					
					97 & 11.24 & 2.23 &	1.89 &	5.50 & 1.99 & 18.23 \\ 
					98 & 10.49 & 4.64 &	1.25 &	14.59 &	7.01 &	6.45 \\ 
					99 & 27.84 & 4.77 &	7.78 &	14.93 &	3.98 &	16.30 \\ 
					100 & 10.48 & 4.28 & 3.07 &	3.28 &	1.14 &	2.29 \\ 
					101 & 1.46 & 2.77 &	0.60 &	3.10 &	2.57 &	8.72 \\ 
					102 & 6.34 & 6.97 &	32.60 &	6.36 &	2.70 &	20.34 \\ 
					103 & 31.52 & 3.03 & 3.42 &	16.72 &	1.49 &	9.87 \\ 
					104* & 5.15 & 2.46 & 2.26 &	4.14 &	1.86 &	4.52 \\ 
					105* & 3.47 & 1.65 & 1.09 &	36.34 &	1.73 &	1.87 \\ 
					106 & 23.62 & 8.48 &  4.92 & 13.53 & 7.17 &	4.01 \\ 
					107 & 33.93 & 10.58 & 9.87 & 61.34 & 12.83 & 17.35 \\ 
					108* & 8.41 & 0.55 & 0.63 &	18.10 &	2.63 &	12.23 \\ 
					109* & 2.37 & 4.20 & 2.52 &	3.67 &	0.81 &	2.84 \\ 
					110* & 2.31 & 5.31 & 0.60 &	10.17 &	2.49 &	6.58 \\ 
					111 & 21.07 & 2.83 & 0.56 &	14.96 &	2.90 &	4.01 \\ 
					112 & 8.28 & 2.65 &	0.38 &	6.80 &	2.51 &	6.06 \\ 
					113* & 7.09 & 6.40 & 33.30 & 62.10 & 6.70 &	175.26 \\ 
					114 & 16.49 & 3.59 & 0.86 &	11.06 &	1.29 &	1.73 \\ 
					115 & 6.90 & 11.13 & 1.70 & 28.72 &	11.68 &	0.80 \\ 
					116 & 16.61 & 1.60 & 11.74 & 12.66 & 1.26 &	16.99 \\ 
					117* & 66.65 &  3.93 & 24.50 & 59.62 & 4.27 & 17.83 \\ 
					118 & 17.09 & 3.08 & 10.92 & 6.30 &	0.98 &	7.28 \\ 
					119 & 2.25 & 3.47 &	10.35 &	2.99 &	1.69 &	9.29 \\ 
					120 & 7.82 & 1.81 &	6.17 &	6.61 &	1.83 &	3.14 \\ 
					121 & 3.01 & 6.54 &	2.01 &	17.14 &	2.84 &	10.32 \\ 
					122 & 95.55 & 4.59 & 0.56 &	31.61 &	3.10 &	24.51 \\ 
					123* & 8.68 & 56.12 & 5.93 & 84.04 & 60.46 & 2.50 \\ 
					124 & 314.97 & 20.82 &	8.30 & 220.67 & 13.09 & 6.22 \\ 
					125 & 14.69 & 3.44 & 5.16 &	14.44 &	3.35 &	4.52 \\ 
					126 & 1.98 & 4.34 &	0.89 &	25.11 &	9.73 &	4.38 \\ 
					127* & 27.89 & 4.23 & 10.20 & 17.13 & 3.38 & 8.68 \\ 
					128 & 63.11 & 3.40 & 18.11 & 57.82 & 2.83 &	13.88 \\ 
					129* & 8.10 & 8.64 & 8.01 & 17.45 &	4.46 &	4.86 \\ 
					130* & 35.51 & 2.26 & 3.64 & 39.95 & 2.62 &	3.29 \\ 

					\hline
					
				\end{tabular}
			\end{center}
			
		\end{table}
		
	\end{landscape}

	\begin{landscape} 
		
		\begin{table}
			\caption{continued}
			\begin{center}
				\begin{tabular}{|c|ccc|ccc|c|}
					
					\hline
					\hline
					CME        & \multicolumn{3}{c|}{Using BG1} &  \multicolumn{3}{c|}{Using BG2} \\ 
					event      &  $C_{\rm x}$  & $C_{\rm p}$    &  $C_{\rm r}$ &  $C_{\rm x}$  & $C_{\rm p}$    &  $C_{\rm r}$    \\
					number    &    &    &    &   &   &    \\
					\hline

					131* & 3.88 & 1.92 & 4.45 &	45.50 &	2.71 &	21.16 \\ 
					132 & 2.64 & 9.18 &	0.68 &	6.94 &	2.07 &	4.54 \\ 
					133 & 10.42 & 4.57 & 0.80 &	3.57 &	1.68 & 3.15 \\ 
					134 & 60.52 & 7.87 & 3.75 &	60.83 &	8.14 &	4.25 \\ 
					135 & 70.57 & 11.85 & 6.38 & 60.93 & 10.70 & 30.87 \\ 
					136 & 26.59 & 1.30 & 1.46 &	11.22 &	0.47 & 2.26 \\ 
					137 & 3.24 & 4.63 &	2.49 &	4.56 &	1.05 &	5.80 \\ 
					138 & 2.72 & 6.16 &	0.94 &	11.38 &	6.84 &	4.85 \\ 
					139 & 19.44 & 3.57 & 6.46 & 17.86 & 4.86 &	4.74 \\ 
					140 & 167.01 & 6.27 & 2.72 & 17.32 & 0.33 &	4.11 \\ 
					141 & 24.61 & 7.93 & 8.90 &	18.73 &	6.53 &	9.87 \\ 
					142 & 12.24 & 4.65 & 2.51 &	17.41 &	5.81 &	1.91 \\ 
					143 & 1.68 & 4.68 &	1.88 &	13.04 &	4.24 &	5.25 \\ 
					144 & 2.58 & 2.19 &	0.02 &	0.84 &	0.72 &	-0.05 \\ 
					145 & 11.09 & 5.61 & 1.37 &	8.38 &	5.30 &	6.93 \\ 
					146 & 9.25 & 4.18 &	7.89 &	3.90 &	1.55 &	3.03 \\ 
					147 & 69.95 & 12.32 & 54.23 & 17.57 & 4.58 & 8.12 \\ 
					148 & 16.97 & 7.82 & 8.60 &	18.11 &	14.99 &	6.20 \\ 
					149 & 8.60 & 2.94 &	1.68 &	4.33 &	2.82 &	2.16 \\ 
					150 & 11.09 & 5.12 & 6.40 &	7.57 &	2.83 &	4.65 \\ 
					151 & 6.06 & 8.26 &	1.38 & 3.61 & 6.24 & 2.08 \\ 
					152 & 10.81 & 5.11 & 2.08 &	12.09 &	9.06 & 5.98 \\ 

					\hline
					
				\end{tabular}
				\label{S - Table B}
			\end{center}
			
		\end{table}
		
	\end{landscape}

\end{appendix}
\end{document}